\documentclass[twocolumn,superscriptaddress,nopacs,aps,pra]{revtex4-2}
\usepackage[english]{babel}
\usepackage[utf8]{inputenc}
\usepackage{soul}
\usepackage{mathtools}
\usepackage{physics}
\usepackage[dvipsnames]{xcolor}
\usepackage{graphicx}
\usepackage{array}
\usepackage{tabularx}
\usepackage[T1]{fontenc}
\usepackage[pdftex, pdftitle={Article}, pdfauthor={Author}]{hyperref}
\usepackage[normalem]{ulem}

\begin{document}
\title{Tailoring bistability in optical tweezers with vortex beams and spherical aberration}

\author{Arthur L. da Fonseca}
\affiliation{Instituto de F\'{\i}sica, Universidade Federal do Rio de Janeiro \\ Caixa Postal 68528,   Rio de Janeiro,  Rio de Janeiro, 21941-972, Brazil}
\affiliation{CENABIO - Centro Nacional de Biologia Estrutural e Bioimagem, Universidade Federal do Rio de Janeiro,
Rio de Janeiro, Rio de Janeiro, 21941-902, Brazil}

\author{Kain\~a Diniz}
\affiliation{Instituto de F\'{\i}sica, Universidade Federal do Rio de Janeiro \\ Caixa Postal 68528,   Rio de Janeiro,  Rio de Janeiro, 21941-972, Brazil}
\affiliation{CENABIO - Centro Nacional de Biologia Estrutural e Bioimagem, Universidade Federal do Rio de Janeiro,
Rio de Janeiro, Rio de Janeiro, 21941-902, Brazil}

\author{Paula B. Monteiro}
\affiliation{Instituto Federal de Santa Catarina, Campus Florianopolis, Av. Mauro Ramos, 950,  Florian\'opolis, Santa Catarina, 88020-300, Brazil }

\author{Lu\'{\i}s B. Pires}
\affiliation{Université de Strasbourg, CNRS, Institut de Science et d’Ingénierie Supramoléculaires, UMR 7006, F-67000 Strasbourg, France}

\author{Guilherme T. Moura}
\affiliation{Instituto de F\'{\i}sica, Universidade Federal do Rio de Janeiro \\ Caixa Postal 68528,   Rio de Janeiro,  Rio de Janeiro, 21941-972, Brazil}
\affiliation{CENABIO - Centro Nacional de Biologia Estrutural e Bioimagem, Universidade Federal do Rio de Janeiro,
Rio de Janeiro, Rio de Janeiro, 21941-902, Brazil}

\author{Mateus Borges}
\affiliation{Instituto de F\'{\i}sica, Universidade Federal do Rio de Janeiro \\ Caixa Postal 68528,   Rio de Janeiro,  Rio de Janeiro, 21941-972, Brazil}
\affiliation{CENABIO - Centro Nacional de Biologia Estrutural e Bioimagem, Universidade Federal do Rio de Janeiro,
Rio de Janeiro, Rio de Janeiro, 21941-902, Brazil}

\author{Rafael S. Dutra}
\affiliation{LISComp-IFRJ, Instituto Federal de Educa\c c\~ao, Ci\^encia e Tecnologia, Rua Sebast\~ao de Lacerda, Paracambi, Rio de Janeiro, 26600-000, Brasil}

\author{Diney S. Ether Jr}
\affiliation{Instituto de F\'{\i}sica, Universidade Federal do Rio de Janeiro \\ Caixa Postal 68528,   Rio de Janeiro,  Rio de Janeiro, 21941-972, Brazil}
\affiliation{CENABIO - Centro Nacional de Biologia Estrutural e Bioimagem, Universidade Federal do Rio de Janeiro,
Rio de Janeiro, Rio de Janeiro, 21941-902, Brazil}

\author{Nathan B. Viana}
\affiliation{Instituto de F\'{\i}sica, Universidade Federal do Rio de Janeiro \\ Caixa Postal 68528,   Rio de Janeiro,  Rio de Janeiro, 21941-972, Brazil}
\affiliation{CENABIO - Centro Nacional de Biologia Estrutural e Bioimagem, Universidade Federal do Rio de Janeiro,
Rio de Janeiro, Rio de Janeiro, 21941-902, Brazil}

\author{Paulo A. Maia Neto}
\email{pamn@if.ufrj.br}
\affiliation{Instituto de F\'{\i}sica, Universidade Federal do Rio de Janeiro \\ Caixa Postal 68528,   Rio de Janeiro,  Rio de Janeiro, 21941-972, Brazil}
\affiliation{CENABIO - Centro Nacional de Biologia Estrutural e Bioimagem, Universidade Federal do Rio de Janeiro,
Rio de Janeiro, Rio de Janeiro, 21941-902, Brazil}

\date{\today}

\begin{abstract}
We demonstrate a bistable optical trap by tightly focusing a vortex laser beam. 
The optical potential has the form of a Mexican hat with an additional minimum at the center.
The bistable trapping corresponds to a non-equilibrium steady state (NESS), where the microsphere 
continually hops, due to thermal activation, between an axial equilibrium state and an orbital 
state driven by the optical torque.
We develop a theoretical model for the optical force field, based entirely on experimentally
accessible parameters, combining a Debye-type non-paraxial description of the focused vortex 
beam with Mie scattering by the microsphere. The theoretical prediction that the microsphere 
and the annular laser focal spot should have comparable sizes is confirmed experimentally by 
taking different values for the topological charge of the vortex beam. Spherical aberration 
introduced by refraction at the interface between the glass slide and the sample is taken into 
account and allows to fine tune between axial, bistable and orbital states as the sample is shifted 
with respect to the objective focal plane. We find an overall agreement between theory and 
experiment for a rather broad range of topological charges. Our results open the way for 
applications in stochastic thermodynamics as it establishes a new control parameter, the 
height of the objective focal plane with respect to the glass slide, that allows to shape 
the optical force field in real time and in a controllable way. 
\end{abstract}


\maketitle

\section{Introduction}\label{intro}

Optical tweezers~\cite{Ashkin1986,Ashkin2006,Gennerich2016}
with structured light beams~\cite{Forbes2021} allow for
a vast range of applications in optical micromanipulation \cite{Rubinsztein-Dunlop2017,Shen2019,Yang2021,Li2021b,Bobkova2021,Spreeuw2022}. 
Photons in a vortex beams carry orbital angular momentum \cite{Allen1992}, which can be exchanged with the trapped particle as an optical 
torque~\cite{He1995,Simpson1997}. 
As a result of spin-orbit coupling~\cite{Bliokh2011,Bliokh2015,Kotlyar2020}, strong focusing of a circularly polarized vortex beam produces 
an annular focal spot whose properties depend on the relative sign 
between the orbital and spin angular momenta~\cite{Ganic2003,Bokor2005,Iketaki2007,Monteiro2009,Philbin2018,Kotlyar2019}. 
In the standard optical tweezers setup, such annular spot 
provides for two very different trapping conditions. 
Particles smaller than the ring of maximum energy density 
resolve the spatial energy variation and, as a consequence, 
move along a circular orbit around the optical axis~\cite{Volke-Sepulveda2002,Curtis2003}. On the other hand, 
larger particles are expected to be trapped on a stable on-axis position~\cite{Ng2010,Zhou2017}. 

In this paper, we demonstrate, both theoretically and experimentally, 
that a bistable trapping is achieved as the orbital and axial states co-exist in the intermediate size range. 
The trapping potential has the form of a Mexican hat with an additional minimum at the center. 
The onset of bistability, as well as the transition from axial to orbital trapping, can be controlled by 
adjusting the focal height with respect to the glass slide at the bottom of our sample. Indeed, as the 
spherical aberration phase introduced by refraction at the glass-water interface is proportional to the focal 
height~\cite{Torok95}, we are able to switch from axial to bistable and then to orbital trapping by displacing 
the sample with the help of a piezoelectric nanopositioning system.

Thus, our system combines paradigmatic models of stochastic thermodynamics~\cite{Seifert2012,Ciliberto2017} into a single platform disposing of a 
tunable parameter, the focal height, allowing to explore different trapping regimes.  
Indeed, by increasing the focal height, we drive the Brownian particle from the  equilibrium state in a harmonic potential to the regime with 
two distinct mesostates \cite{Seifert2019} characterised by different conformational free energies \cite{Roldan2014}.
Then, by further increasing the focal height, we implement 
a paradigmatic non-equilibrium steady state (NESS)~\cite{Seifert2012},
in which a colloidal particle is driven along a circular orbit by the non-conservative optical force component associated to the laser beam angular momentum.
To explain our experimental results, we extend the 
Mie-Debye spherical aberration (MDSA) theory of optical tweezers~\cite{Neto2000,Mazolli2003,Viana2007,Dutra2007} by considering 
a vortex beam at the objective entrance port. 
The paraxial approximation is taken only at the entrance port, and the 
non-paraxial tightly focused trapping beam arises as  
 a vector interference of spatial Fourier components~\cite{Richards1959}.
Such realistic description allows us to analyze in detail how the optical force field changes as the spherical aberration increasingly degrades the focal spot. 

The paper is organized as follows. The experimental setup and the theoretical formalism are presented in Secs.~\ref{sec:experiment} and 
\ref{sec:theory}, respectively. 
Bistability is first discussed in the simpler scenario of 
 ideal aberration-free trapping beams in Sec.~\ref{sec:Trapping regimes}, while 
experimental and theoretical results accounting for spherical aberration are compared in Sec.~\ref{subsec:sphericalaberration}.
 Sec.~\ref{sec:conclusion} is devoted to concluding remarks.

\section{Experimental setup}
\label{sec:experiment}

\begin{figure}[ht]
    \includegraphics[scale=.12]{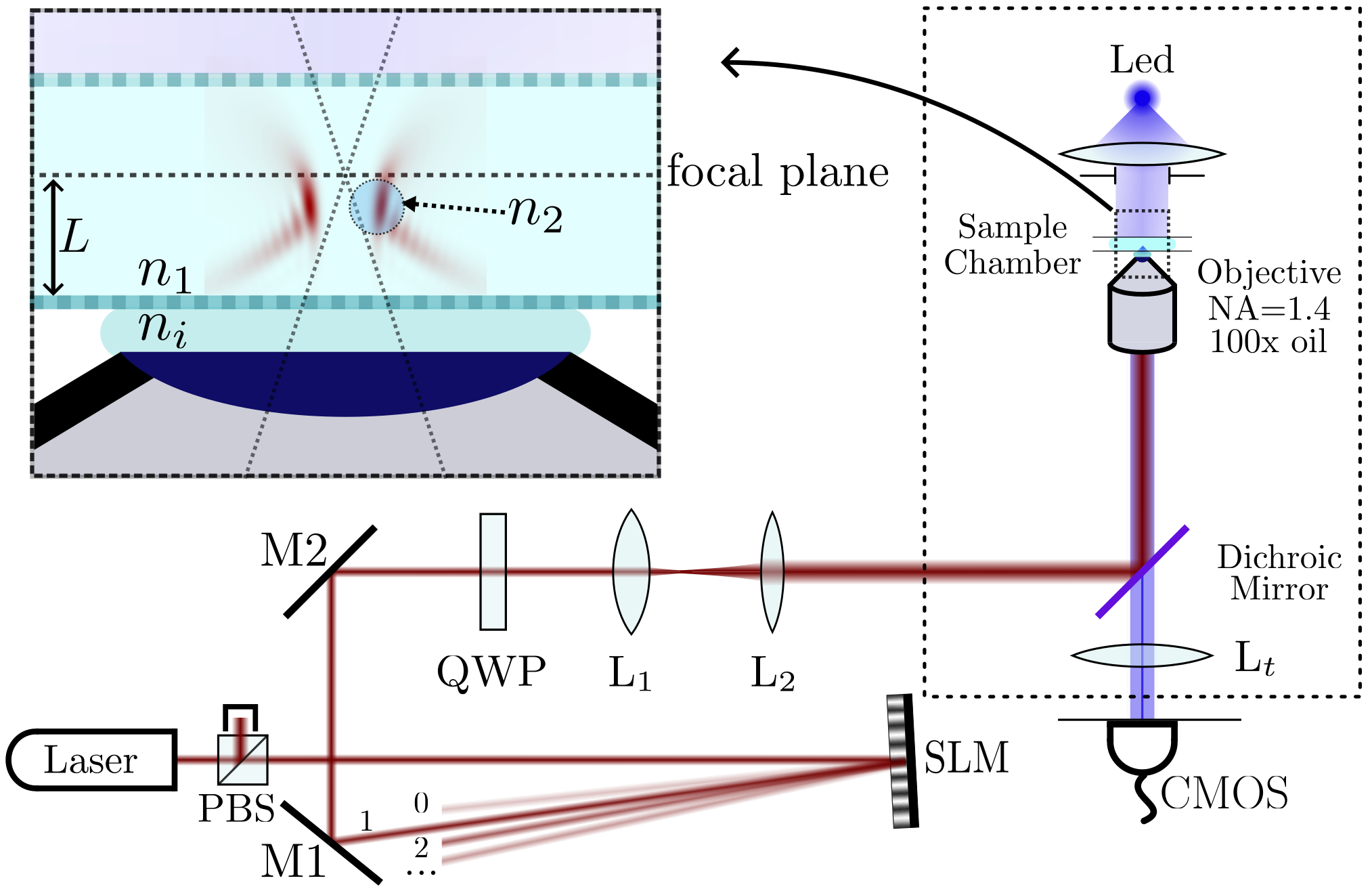}
    \caption{Illustration of the experimental setup. The laser beam  goes through a polarizing beam spliter (PBS) and is directed to the spatial 
    light modulator (SLM). The first order of diffraction  propagates through a quarter-wave plate (QWP) and, for values of $\ell$ up to $5$, through 
    a beam expander (lenses L1 and L2) towards the microscope (dotted frame). Inside the microscope, the beam is reflected by a dichroic mirror (DM) 
    and then focused by an oil-immersion objective. 
    The inset shows a magnified view of the sample region with the glass slide lying at its bottom.  The amount of spherical aberration introduced by 
    refraction at the interface between the glass slide and the sample is controlled by changing the height $L$ of the objective focal plane with respect 
    to the slide, whose position is controlled by a piezoelectric nano-positioning stage (not shown). The resulting nonparaxial focused beam is indicated 
    by the density plot of the electric energy density for $L=10\,\mu{\rm m}$ (red).
    }
    \label{FigExp}
\end{figure}

The generation of structured light beams has been extensively discussed~\cite{rosales2017shape}. 
Here we use a spatial light modulator (SLM) to synthesize a vortex beam with topological charge $\ell.$ 
Our setup is depicted in Fig.~\ref{FigExp}. We steer a horizontally polarized ${\rm TEM}_{00}$ laser beam (IPG photonics, model YLR-5-1064LP) with 
wavelength $\lambda_0 = 1064\, {\rm nm}$ onto the SLM (Holoeye Photonics AG Pluto), in which we display an overlap between the vortex phase and a 
linear ramp, producing several orders of diffraction. 

As shown in Fig.~\ref{FigExp}, we propagate the first order of diffraction through a quarter-wave plate (QWP) to produce left-handed 
circular polarization and expand its beam waist $w_0$ so that the annular vortex beam slightly overfills the objective entrance port. 
The overfilling of the objective entrance port of radius $R_p=(2.80 \pm 0.05)\, {\rm mm}$ is such that the radius of maximum 
intensity $ r_\ell = w_0 \sqrt{\frac{|\ell|}{2}}$ is of the order of $R_p.$  The reason for selecting such filling condition as well 
as the values of $w_0$ and $ r_\ell$ are discussed in detail in Appendix \ref{sec:beam waist}.
We check the transverse profile of the vortex beam after expansion with the help of a scanning-slit beam profiler (ThorLabs BP209-VIS/M). 

After reflection by a dichroic mirror, the vortex beam is strongly focused  by the oil-immersion microscope objective (Nikon PLAN APO, 100x, NA = 1.4). 
The sample chamber is contained by an O-ring on top of the glass slide and filled with a suspension of polystyrene microspheres (Polysciences, Warrington, PA) in water.
The entire system is displaced vertically by a piezoelectric nano-positioning stage (Digital Piezo Controller E-710, Physik Instrumente), allowing us to change 
the distance $L$ between the glass slide and the objective focal plane as depicted in the inset of Fig.~\ref{FigExp}. 
Since the spherical aberration phase introduced by refraction at the interface between the glass slide and the aqueous medium is proportional to $L$ \cite{Torok95}, we can 
tailor different trapping regimes by fine tuning such distance. In order to provide room for trapping with small and moderate values of $L,$ we displace the diffraction 
focus upwards with respect to the objective focal plane by allowing the vortex beam to develop a finite curvature as it propagates towards the back aperture of the objective. 
According to the displacement theorem~\cite{Born2019}, the resulting curvature of field leads to a global shift of the laser focal spot without changing the amount of spherical aberration.

The Köhler illumination by a LED source (wavelength $470\,{\rm nm}$) is also depicted in Fig.~\ref{FigExp}. Light scattered by the microspheres is collected by the objective 
and goes through the dichroic mirror and the microscope tube lens L$_t$.
The resulting images are recorded by a CMOS camera (Hamamatsu Orca-Flash 2.8 C11440-10C) for data analysis.

\section{MDSA theory of optical trapping with vortex beams}
\label{sec:theory}

MDSA theory~\cite{Neto2000,Mazolli2003,Viana2007, Dutra2007} combines a nonparaxial Debye-type model of a strongly focused beam~\cite{Richards1959} with Mie scattering by the trapped microsphere. 
When astigmatism is included, good agreement with experimental data for the trap stiffness~\cite{Dutra2012,Dutra2014} and the vorticity at the focal point~\cite{Diniz2019} is found, with no fitting.  
Here, we extend MDSA theory in order to account for focusing of vortex  beams. For simplicity, we neglect astigmatism and model the paraxial beam entering the the objective back aperture as a 
circularly-polarized Laguerre-Gaussian LG$_{0\ell}$ mode with radial order $p=0$. In cylindrical coordinates, the corresponding electric field reads
\begin{equation}
    \mathbf{E}_{\rm p}(\rho, \phi,z) = E_0 \left(\frac{\sqrt{2}\rho}{w_0}\right)^{|\ell|} e^{-\frac{\rho^2}{w_0^2}} e^{i\ell\phi} e^{i k_0z}\boldsymbol{\hat\epsilon}_{\sigma}.
    \label{paraxial_LG}
\end{equation}
Here, $k_0=2\pi/\lambda_0$ is the laser wavenumber and  $\boldsymbol{\hat\epsilon}_{\sigma}=(\mathbf{\hat x}+i\sigma\mathbf{\hat y})/\sqrt{2}$ are the unit vectors for left-handed 
($\sigma= 1$) and right-handed ($\sigma= -1$) circular polarizations.
More general vortex beams and polarizations can also be analysed by using the method outlined below. 

The strongly focused non-paraxial beam after the objective of numerical aperture $\mbox{NA}$ 
(obeying the sine condition) is obtained from (\ref{paraxial_LG}) as a Debye-type (Fourier) superposition of plane waves \cite{Richards1959} with wavevectors ${\bf k}(\theta,\varphi)$ spanning the 
angular sector defined by the conditions $ 0\le \varphi < 2\pi$ and $0\le \theta\le \theta_0= \sin^{-1}\left(\frac{{\rm NA}}{n_i}\right),$ 
where $n_i$ is the refractive index of glass. All Fourier components satisfy $|{\bf k}(\theta,\varphi)|=n_ik_0.$

The focused beam is then further refracted at the interface between the glass slide and the sample region filled with water shown in the inset of Fig.~\ref{FigExp}. 
In a typical oil-immersion objective, 
such refraction has an important effect on the optical force as it degrades the focal region by the introduction of spherical aberration. For each Fourier component, 
the resulting spherical aberration phase scales with the distance $L$ between the objective focal plane and the slide~\cite{Viana2007} shown in the inset of Fig.~\ref{FigExp}:  
\begin{equation}
    \Phi(\theta)=kL\left(-\frac{\cos\theta}{N_a} + N_a\,\cos\theta_1\right),
    \label{phi_SA}
\end{equation}
where $\theta_1$ is the refraction angle in the sample filled with water and $N_a=n_1/n_i$ is the relative refractive index of  water with respect to the glass medium. 

For high-NA objectives, the part of the angular spectrum corresponding to $\theta>\sin^{-1}\left(N_a\right)$ gives rise to evanescent waves in the sample region. We 
assume that the trapped particle is a few wavelengths away from the glass slide at the bottom of the sample region, allowing us to neglect the contribution of the 
evanescent sector as well as the effect of optical reverberation between particle and glass slide \cite{Dutra2016}. We discard the contribution from the evanescent 
sector by taking $ \theta_0=\sin^{-1}\left(N_a\right)$ when ${\rm NA}>n_1.$

We compute the Mie scattered field for each component of the angular spectrum of the nonparaxial incident field with the help of 
the Wigner rotation matrix elements~\cite{Edmonds1957} $d^j_{m,m'}(\theta)$ allowing us to consider all directions of incidence contained in the spectrum. 
Finally, the optical force ${\bf F}$ is derived from the Maxwell stress tensor. As the former is proportional to the laser beam power $P$ at the sample, we define the 
dimensionless force~\cite{Ashkin1992}
\begin{equation}
    \begin{split}
    \mathbf{Q} = \frac{\mathbf{F}}{n_1\, P/c}.
    \label{vecQ_def}
    \end{split}
\end{equation}
where $c$ is the speed of light in vacuum. 
The resulting optical force is the sum of two contributions~\cite{Neto2000}: the extinction term $\mathbf{Q}_e$ represents the 
rate of linear momentum removal from the incident field. Part of this momentum is carried away by the scattered field at a rate $-\mathbf{Q}_s.$
The rate of momentum transferred to the particle is then $\mathbf{Q}=\mathbf{Q}_e+\mathbf{Q}_s.$ 
The cylindrical components are written as partial-wave series of the form 
\[
\sum_{jm} = \sum_{j=1}^{\infty}\sum_{m=-j}^j.
\]
The axial extinction contribution reads
\begin{equation}
    \begin{split}
    Q_{ez} = \frac{2(2\gamma^2)^{|\ell|+1}}{|\ell|! \, A_{\ell} \, N_a }\Re\sum_{jm}(2j+1)(a_j + b_j)\big(G_{j,m}G'^*_{j,m}\big).
    \label{Qez}
    \end{split}
\end{equation}
Here, $a_j$ and $b_j$ are the Mie coefficients for electric and magnetic multipoles \cite{Bohren1998}, respectively,
\begin{equation}
    \begin{split}
    A_{\ell} = \frac{8(2\gamma^2)^{|\ell|+1}}{|\ell|!} \int^{\sin{\theta_0}}_0 ds \,s^{2|\ell|+1} \exp(-2(\gamma s)^2)\\
    \times \frac{ \sqrt{(1-s^2)(N_a^2-s^2)} }{ (\sqrt{1-s^2}+\sqrt{N_a^2-s^2})^2 }
    \label{Al_SA}
    \end{split}
\end{equation}
is the fraction of the trapping beam power that fills the objective entrance port of radius $R_p$ and is refracted into the sample, 
and $\gamma=f/w_0$ is the ratio between the objective focal length and the beam waist at the entrance port. 

The multipole coefficients appearing in (\ref{Qez}) are given by
\begin{eqnarray}
    G_{jm} &=& \int^{\theta_0}_0 d\theta (\sin\theta)^{1+|\ell|}\sqrt{\cos\theta}\,
    T(\theta)\, e^{-\gamma^2 \sin^2\theta} \label{G}  \\
    &\times& d^j_{m,1}(\theta_1)\, J_{m-1-\ell}(k\rho\sin\theta_1)\,e^{i \Phi(\theta)}\,e^{i n_1k_0 \cos\theta_1 z}\nonumber
\end{eqnarray}
\begin{eqnarray}
    G'_{jm} &=& \int^{\theta_0}_0 d\theta (\sin\theta)^{1+|\ell|}\sqrt{\cos\theta}
    \,\cos\theta_1\,T(\theta)\, e^{-\gamma^2 \sin^2\theta} \nonumber \\
    &\times& d^j_{m,1}(\theta_1)\, J_{m-1-\ell}(k\rho\sin\theta_1)\,e^{i \Phi(\theta)}\,e^{i n_1k_0 \cos\theta_1 z},
    \label{Gc}
\end{eqnarray}
where $J_{m}$ are the cylindrical Bessel functions of integer order~\cite{DLMF-10}.
The coefficient 
\begin{equation}
    T(\theta) = \frac{2\cos\theta}{\cos\theta + N_a\cos\theta_1}
    \label{tperp}
\end{equation}
is the Fresnel transmission amplitude for the glass-water interface. 

The remaining cylindrical components of $ {\bf Q}_e$ as well as the components of $ {\bf Q}_s$ are written in a similar way. 
Explicit expressions can be found in Appendix~\ref{sec:explicit}.

\section{Results}
\label{sec:results}

\subsection{Trapping states for aberration-free systems}
\label{sec:Trapping regimes}
For clarity, we first discuss the case of an aplanatic focused beam, which 
in principle can be implemented with a water-immersion objective. 
We then take $N_a=1$ leading to a vanishing spherical aberration phase $\Phi=0.$

In this case, the electric energy density in the focal region was discussed in Refs.~\cite{Ganic2003,Bokor2005,Iketaki2007,Monteiro2009}. 
For any nonzero topological charge, it has the shape of a ring that depends on the relative sign between $\ell$ and $\sigma.$ 
On the focal plane, the peak electric energy density is at a distance $\tilde{r}_\ell$ from the axis, which was shown to scale linearly 
with $\ell$ for a fixed waist $w_0$ when $\ell\gg 1$~\cite{Curtis2003,Monteiro2009}. The variation of $\tilde{r}_\ell$ with $\ell$ and 
$r_{\ell}/R_p$ (defining how the vortex beam fills the objective entrance port) is discussed in detail in Appendix~\ref{sec:beam waist}. 
For all numerical results presented in the present subsection, we take the value $r_{\ell}/R_p=0.8,$ which provides a diffraction limited 
spot with a relatively small power loss as discussed in Appendix~\ref{sec:beam waist}. 
We also take $n_1=1.332$ and $n_2=1.576$ for the refractive indexes of water and polystyrene, respectively.

Two trapping regimes are expected depending on the comparison between the  microsphere
radius $a$ and the characteristic size $\tilde{r}_\ell$ of the focal spot~\cite{Ng2010,Zhou2017,Yang2021}.
When $a\ll \tilde{r}_\ell$, the particle is trapped near the ring of maximum energy density while being driven by the optical torque  \cite{Volke-Sepulveda2002,Curtis2003}. 
This is in line with the simple Rayleigh picture of an optical force proportional to the gradient of the electric energy density, alongside a non-conservative force component 
that drives the particle around the beam axis. In the opposite limit of radius $a\gg \tilde{r}_\ell,$ the microsphere is trapped on the optical axis as it is too big to 
resolve the spatial variation of the annular focal spot. 

Our Mie-Debye results presented below confirm the existence of these two trapping regimes. More importantly, we find that at intermediate particle sizes, $a\sim \tilde{r}_\ell,$
the two stable trapping states co-exist, with the particle randomly hopping between the axis and the annular focal spot by thermal activation.

\begin{figure}[ht]
    \includegraphics[scale=.48]{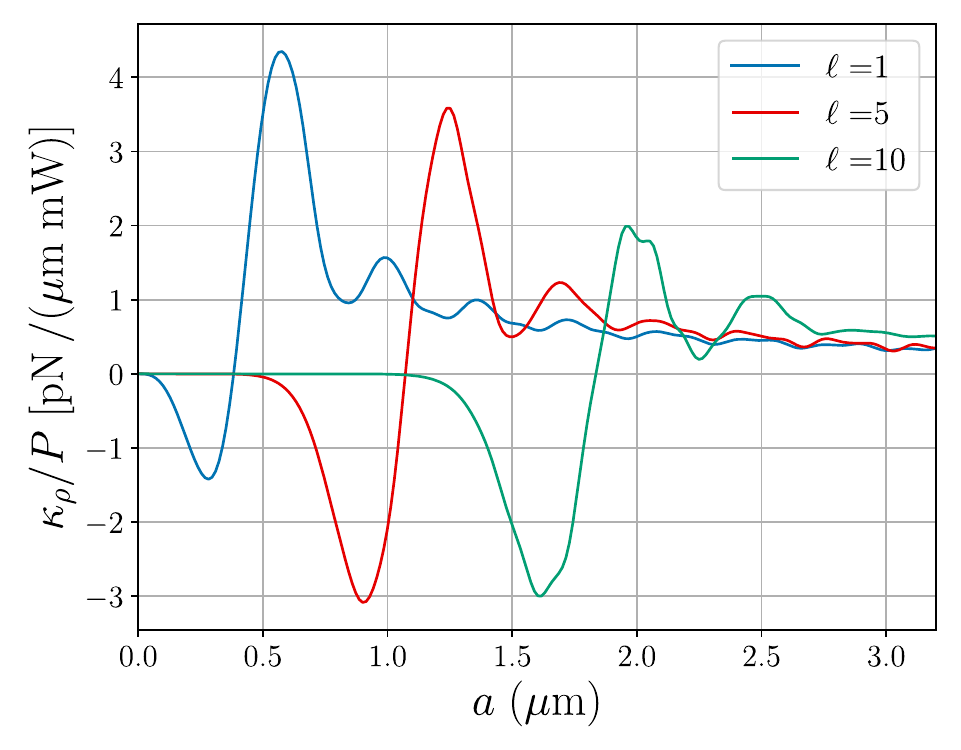}
    
    \caption{Transverse trap stiffness per unit power $\kappa_\rho/P$ as a function of the sphere radius for topological charges $\ell=1$ (blue), $5$ (red) and $10$ (green). We consider
    an aberration-free trapping beam. 
    As the particle size increases, 
    $\kappa_\rho$ changes sign from negative to positive  at the critical radius $R_{\rm on}$.}
    \label{krho_radius}
\end{figure}

\begin{figure}
    \includegraphics[scale=.53]{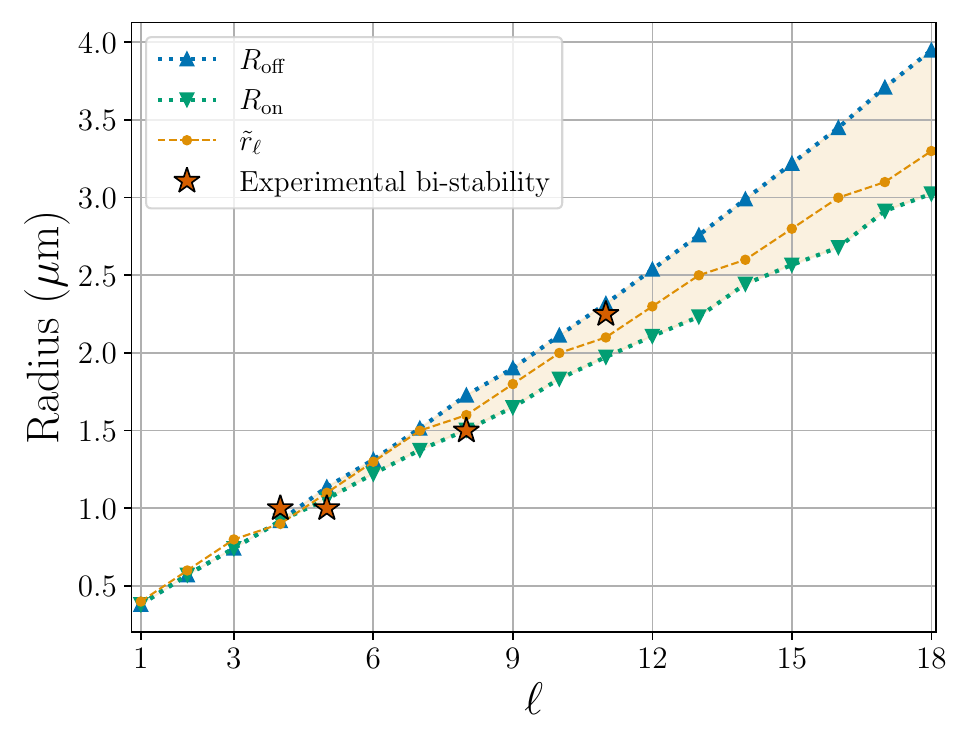}
    
    \caption{For an aberration-free trapping beam,
    the parameter space
    spanned by 
    the microsphere radius and the topological charge $\ell$ is divided into three different regions. For radii smaller than the critical radius $R_{\rm on}(\ell)$ (green), on-axis trapping 
    is excluded and the particle rotates around the optical axis along a circular trajectory. For radii larger than the critical radius $R_{\rm off}(\ell)$ (blue), orbital trapping is excluded 
    and the particle is trapped at a position along the optical axis. On-axis and orbital states co-exist in the (colored) bistable stripe bounded by the plots of $R_{\rm on}(\ell)$ and 
    $R_{\rm off}(\ell).$ Such region corresponds to radii close to the maximum of electric energy density $\tilde{r}_\ell$ (circle). When an oil-immersion objective is employed, the spherical 
    aberration introduced by refraction at the glass slide opens the way to switch between trapping regimes by changing the height $L$ of the objective focal plane. However, we still find an 
    overall qualitative agreement between the simplified aberration-free prediction and experimental realizations of bistability (star) provided that we fine tune the height $L$.}
    \label{phase_diag}
\end{figure}

For on-axis trapping, it is required that the radial trap stiffness satisfies $\kappa_\rho = -(n_1P/c)(\partial Q_\rho/\partial \rho)|_{\rho=0}>0.$ 
We first compute the stable axial position $z_{\rm eq}$ by solving $Q_z(\rho=0,z_{\rm eq})=0$ and then calculate the numerical derivative of the 
function $ Q_\rho(\rho,z_{\rm eq})$ at $\rho=0.$ In Fig.~\ref{krho_radius}, we show the variation of $\kappa_\rho/P$ with microsphere radius for 
different values of $\ell.$ For any positive value of $\ell,$ we find that $\kappa_\rho$ changes its sign from negative to positive at a critical 
sphere radius $R_{\rm on}$ that increases with $\ell.$ Thus, on-axis trapping is excluded for sphere radii smaller than $R_{\rm on}.$ Particles in 
this size range are trapped on the annular region and are driven by the optical torque. In line with the previous qualitative discussion, $R_{\rm on}$ 
is comparable to the focal spot annular radius $\tilde{r}_\ell.$ Indeed, we plot $R_{\rm on}$ (green) and $\tilde{r}_\ell$ (circles) as functions of $\ell$ 
in Fig.~\ref{phase_diag}, showing that $R_{\rm on}$ is slightly smaller than $\tilde{r}_\ell.$ In the region below the line defined by $R_{\rm on}$ in 
Fig.~\ref{phase_diag}, trapping occurs on the annular region only. 

The condition for off-axis trapping defines a second critical radius, $R_{\rm off}.$ When $a>R_{\rm off},$ off-axis trapping is excluded as the only root of $ Q_\rho(\rho,z=0)=0$ is at $\rho=0,$ with 
$Q_\rho(\rho,z=0)<0$ for any $\rho>0.$ In Fig.~\ref{phase_diag}, we plot the variation of $R_{\rm off}$ with $\ell$ (blue). $R_{\rm off}$ is very close to $\tilde{r}_\ell$ for small values of $\ell,$ 
and then becomes increasingly larger than $\tilde{r}_\ell$ as $\ell$ increases. 

\begin{figure}[ht]
    \includegraphics[scale=.53]{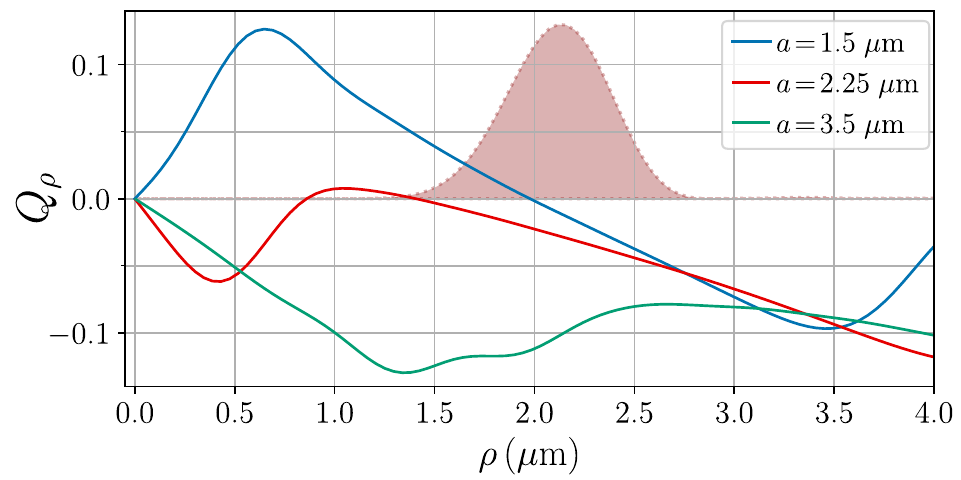}
    
    \caption{Radial optical force component $Q_\rho$ as a function of radial position $\rho$ for $\ell=11$ and radii $a=1.5~\mu{\rm m}$ (blue), $a=2.25~\mu{\rm m}$ (red) and $a=3.5~\mu{\rm m}$ (green). 
    The fill plot indicates the electric energy density variation with distance to the optical axis. Its maximum is located at $\tilde{r}_\ell=2.1~\mu{\rm m}.$}
    \label{Qrho_rho_agua}
\end{figure}

In between the two exclusion zones shown in Fig.~\ref{phase_diag}, corresponding to microsphere radii in the (colored) stripe defined by $R_{\rm on} <a<R_{\rm off},$ both on-axis and off-axis trapping 
are possible, leading to a bistable trap. As an illustration, we plot $Q_{\rho}(\rho,z=0)$ versus $\rho$ in Fig.~\ref{Qrho_rho_agua} for three different microsphere radii: $a=1.5\,\mu{\rm m}$ (blue), 
$2.25\,\mu{\rm m}$ (red) and $3.5\,\mu{\rm m}$ (green). In all cases, we take $\ell=11,$ for which we find $R_{\rm on}=2.0\,\mu{\rm m}$ and $R_{\rm off}=2.3\,\mu{\rm m}.$ Thus, the three radii considered 
in Fig.~\ref{Qrho_rho_agua} illustrate the three trapping regimes defined by the parameter space shown in Fig.~\ref{phase_diag}. For the smallest size, Fig.~\ref{Qrho_rho_agua} shows that the axial 
equilibrium position is unstable, in agreement with the results shown in Fig.~\ref{krho_radius}, whereas the positive root of $Q_{\rho}(\rho)=0$ corresponds to stable equilibrium. For the largest particle, 
the only (stable) equilibrium position is at $\rho=0,$ whereas for the intermediate size two stable equilibria are shown. 

For further insight, we also show in Fig.~\ref{Qrho_rho_agua} the electric energy density as a function of $\rho$ (fill plot). The edge of a microsphere with $a=1.5\,\mu{\rm m}$ is located near the inner 
tail of the electric energy density distribution when its center is aligned along the axis. As it would sit almost entirely on the dark central part of the annular spot, stable on-axis equilibrium is 
indeed not possible in this case. On the other hand, a microsphere with $a=3.5\,\mu{\rm m}$ encompasses the entire bright annulus when placed on-axis, which is consistent with stable on-axis trapping. 
Finally, the intermediate size microsphere ($a=2.25\,\mu{\rm m}$), for which a bistable behavior is predicted, is such that its edge nearly coincides with the energy density peak at $\rho=\tilde{r}_\ell.$
Such discussion indicates that the width of the bistable stripe in the parameter space of Fig.~\ref{phase_diag} scales with the width of the focal spot annular region. As the latter increases with $\ell,$ 
we expect the bistable stripe to become wider as $\ell$ increases, which is indeed in agreement with the results shown in Fig.~\ref{phase_diag}.

\subsection{Tuning the spherical aberration to tailor the trapping states}
\label{subsec:sphericalaberration}

For a typical optical tweezers setup employing an oil-immersion objective, the spherical aberration introduced by refraction of the trapping beam at the glass-water interface modifies the results presented 
in the previous sub-section. As the resulting phase (\ref{phi_SA}) is proportional to the height $L$ of the focal plane with respect to the glass slide, it can be fine tuned by displacing the sample with 
the help of a piezoelectric nano-positioning stage as discussed in Sec.~\ref{sec:experiment}. In the present sub-section, we show that spherical aberration provides a useful tool for switching between 
different trapping states for fixed values of particle size and topological charge. 

Although the results of Fig.~\ref{phase_diag} do not take spherical aberration into account, they still provide a useful guide for achieving bistability with our experimental setup employing an 
oil-immersion objective. Indeed, we are able to implement a thermally-activated bistable trapping by fine tuning the height $L$ when taking particles of radius $a$ illuminated by vortex beams of charge $\ell$ 
close to the bistable stripe shown in Fig.~\ref{phase_diag}, but not otherwise. The stars in Fig.~\ref{phase_diag} indicate the experimental implementations of bistability. The corresponding values of $\ell$ 
and $a$ are shown in Table \ref{table_data}. In all of those cases, we start by trapping the microsphere on the axis  with the focal plane close to the glass slide, and then increase the height $L$ by displacing 
the sample downwards. As $L$ increases, we first switch from axial to bistable trapping, and then from bistable to off-axis orbital motion \cite{SM}.

\begin{table}
\begin{tabularx}{0.48\textwidth}{|| >{\centering\hsize=.24\hsize\linewidth=\hsize}X | >{\centering\hsize=.7\hsize}X | >{\centering\arraybackslash}X | >{\centering\arraybackslash}X | >{\centering\arraybackslash}X ||} 
 \hline
$\ell$ & $a \,(\mu \rm m)$ & $\rho_{\rm{theo}} \,(\mu \rm{m})$ & $\rho_{\rm{exp}} \,(\mu \rm{m})$  & $T_{\rm{exp}}\,(\rm s)$ \\

 \hline\hline
 $4$ & $1.0$   & $-$            & $0.6\pm0.1$ & $0.24\pm0.02$ \\
 \hline
 $5$ & $1.0$   & $0.69\pm0.01$  & $0.7\pm0.2$ & $0.20\pm0.01$ \\ 
 \hline
 $8$ & $1.5$   & $1.07\pm0.01$  & $1.0\pm0.2$ & $0.55\pm0.03$ \\
 \hline
 $11$ & $2.25$ & $1.2\pm0.1$    & $1.9\pm0.3$ & $5\pm1$       \\ [.1ex] 
 \hline
\end{tabularx}
\caption{Experimental values for the radius $\rho_{\rm{exp}}$ and period $T_{\rm{exp}}$ of the orbit when the microsphere is off-axis (see panels (b) and (c) in Fig.~\ref{FigExpData}) in the bistable regime. 
The theoretical values for the orbit's radius $\rho_{\rm{theo}}$ are in good agreement with the experimental data for $\ell=5$ and $\ell=8,$ but for $\ell=4$ only on-axis trapping is predicted by the model. }
\label{table_data}
\end{table}

\begin{figure}
    \includegraphics[scale=.355]{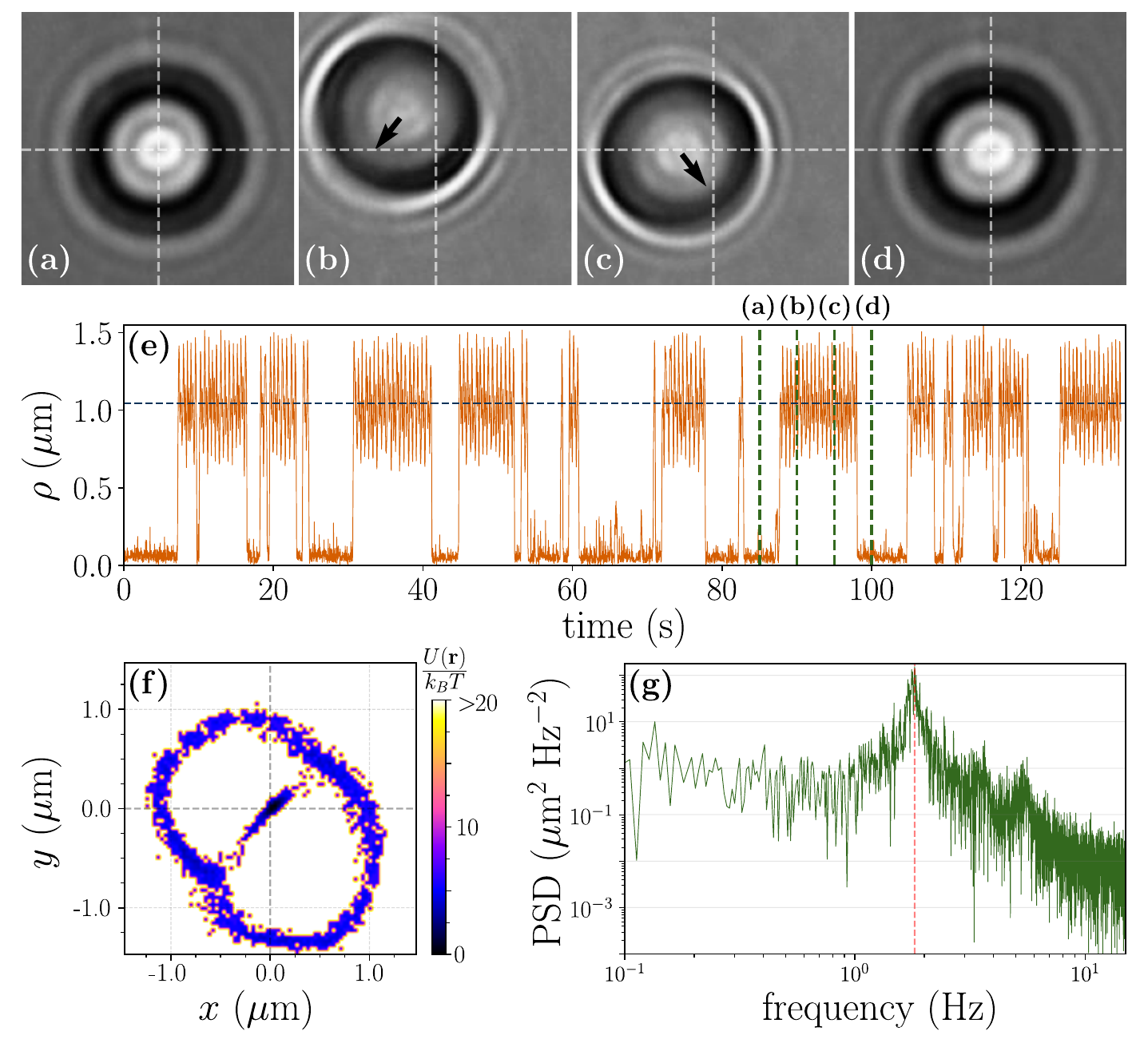}
    
    \caption{Experimental realization of bistability with a polystyrene microsphere of radius $a=1.5\,\mu$m and a vortex beam with $\ell=8$. Panels (a)-(d) show frames of the trapped microsphere at times 
    $t_{(\rm{a})}=85\,{\rm s}$, $t_{(\rm{b})}=90\,{\rm s}$, $t_{(\rm{c})}=95\,{\rm s}$ and $t_{(\rm{d})}=100\,{\rm s}$. When the microsphere hops to an off-axis location (b and c), it circulates around the axis. The black 
    arrows indicate the motion of the microsphere in (b) and (c). (e) Microsphere radial coordinate versus time. The vertical green lines indicate the times corresponding to the frames shown in (a)-(d). The horizontal dashed 
    line represents the mean radial coordinate $\rho_{\rm exp}=(1.04\pm 0.05)\,\mu$m for the off-axis phase. (f) Color map of the energy distribution $U(x, y)/(k_{\rm B}T)$ across the $xy$ plane as derived from the position 
    distribution density $p(x, y)$. (g) Power spectrum density (PSD) of the $x$ coordinate showing a peak at $f=1.82\,{\rm Hz}$, which corresponds to an orbital period $T_{\rm exp}=(0.55\pm 0.03)\,{\rm s}.$ }
    \label{FigExpData}
\end{figure}

A typical bistable trapping obtained at intermediate values of $L$ for $\ell=8$ and $a=1.5\,\mu{\rm m}$ is illustrated by Fig.~\ref{FigExpData}. Panels (a)-(d) show frames of the trapped particle as it hops from 
the axial position to the off-axis orbit and back. The alternation between on and off-axis states over time is presented in more detail in  panel (e), where we plot the microsphere radial position $\rho$ versus time.
We determine the radius of the orbit $\rho_{\rm exp}$ from the average (horizontal dashed line) and the standard error of $\rho$ in the orbital state. The resulting figures for $\ell=8$ as well as for the other 
values of $\ell$  are indicated in Table \ref{table_data}. The instants of time corresponding to panels (a)-(d) are indicated in panel (e) as vertical dashed lines. From the complete trajectory on the $xy$ plane, we 
determine the position distribution density $p(x, y)$ by taking bins of area $\Delta x\Delta y =  1.35\times 10^{-3}\, \rm \mu m^2$. Subsequently, $p(x,y)$ defines the energy distribution 
$U(x, y)/(k_{\rm B}T) = \ln p(0, 0) - \ln p(x, y)$ depicted in panel (f). 

The period of the orbit $T_{\rm exp}$ is obtained from the peak in the power spectrum density (PSD) of the microsphere $x$ coordinate shown in Fig.~\ref{FigExpData}(g).  A very similar result is found for the PSD of 
the $y$ coordinate, as well as for PSDs of the $x$ and $y$ coordinates found for the other values of $\ell$ (not shown). The resulting values of $T_{\rm exp}$ are shown in Table~\ref{table_data}, with the errors bars 
derived from the half width at half maximum. 

In all cases, we find, both theoretically and experimentally, that the sense of rotation coincides with the sign of $\ell.$ This is also the case of a previous experiment with a vortex beam focused into an aqueous 
solution~\cite{Curtis2003}, but a negative optical torque was predicted for the orbital motion in air~\cite{Li2021}. A negative torque was also demonstrated for a particle trapped on-axis by a circularly-polarized 
Gaussian beam~\cite{Diniz2019}.

\begin{figure}[h]
    \includegraphics[scale=.4]{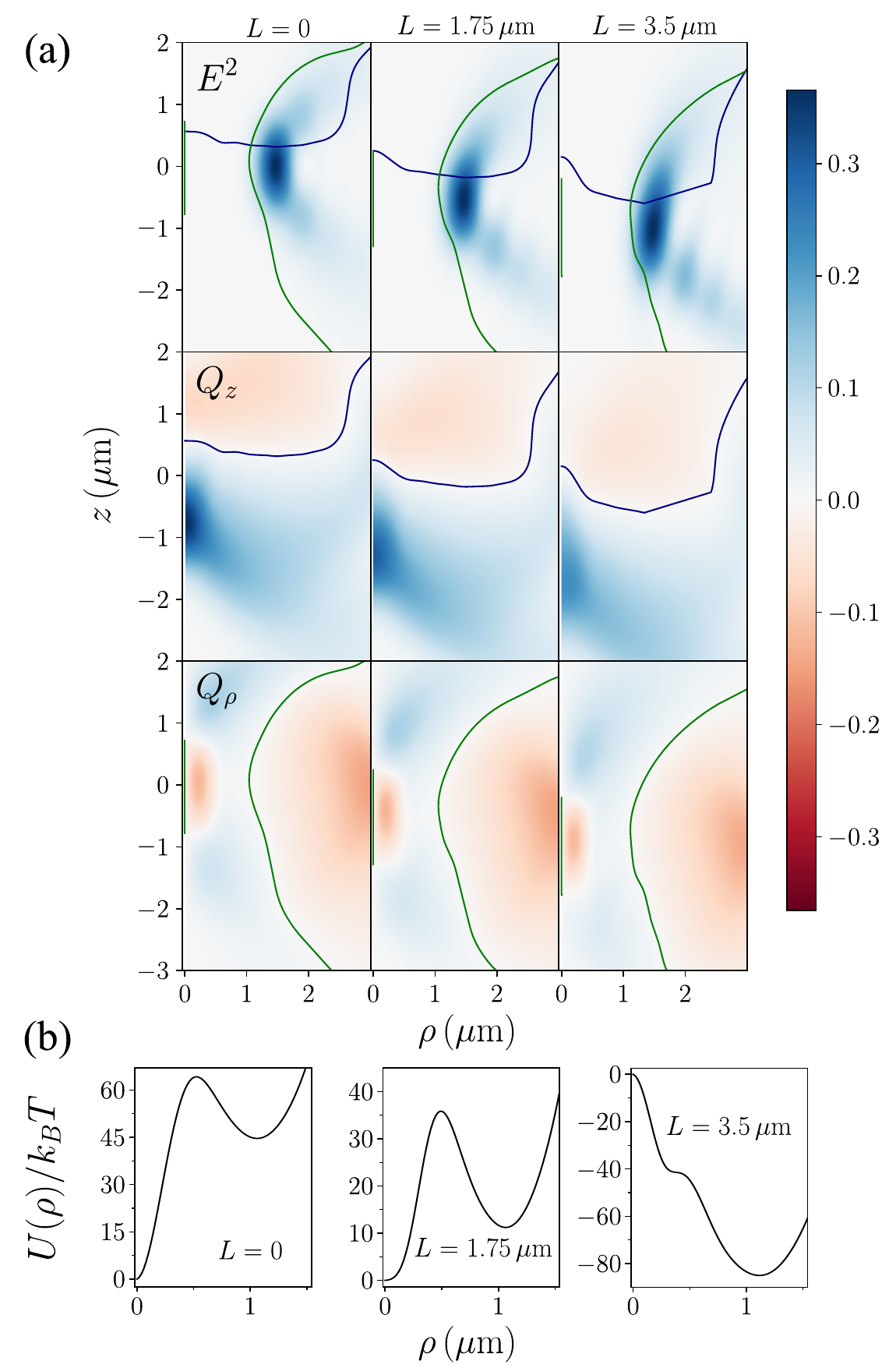}
    \caption{Variation of the optical force field with spherical aberration. 
    The distance $L$ between the objective focal plane (for paraxial rays) and the glass slide is increased from left to right, thus enhancing the spherical aberration introduced by refraction at the interface between the 
    slide and the sample. A vortex beam with  $\ell=8$ and lefthanded circular polarization ($\sigma=+1$) is focused by an oil-immersion objective so as to trap a polystyrene microsphere with radius $a=1.5\,\mu{\rm m}.$ 
    (a) From top to bottom, density plots representing the electric energy density $E^2$, and the axial and radial force components, $Q_z$ and $Q_\rho,$ respectively, on a meridional plane. The force components are normalized 
    by Eq.~(\ref{vecQ_def}) and their values are indicated by the color bar. The blue and green lines indicate the roots of $Q_z=0$ and $Q_{\rho}=0,$ respectively, that lead to stable 3D trapping when they intersect. (b) Optical 
    potential (in units of the thermal energy $k_BT$) versus radial distance $\rho.$ }
    \label{Mosaic}
\end{figure}

To understand in detail how spherical aberration controls the onset of bistability, we plot the electric energy density and optical force components for different values of $L$ in Fig.~\ref{Mosaic}(a), again for 
$\ell=8$ and  $a=1.5\,\mu{\rm m}.$ The columns correspond to the three positions of the objective focal plane with respect to the glass slide employed in the experiment, with the leftmost one depicting trapping 
closer to the slide. From left to right, we displace the sample downwards by steps of $d = 2\,\mu{\rm m},$ which corresponds to a variation of the objective focal plane height of $\Delta L=N_ad=1.75\,\mu{\rm m}.$     
Experimentally, we employ a slightly diverging beam at the objective entrance port, so as to shift the laser focal spot to a position above the objective focal plane by a few microns thus making room for trapping 
in the sample region. The density plots represent the electric energy density $E^2$ and the axial ($Q_z$) and radial ($Q_{\rho}$) optical force components as functions of the microsphere position in cylindrical 
coordinates. The plane $z=0$ corresponds to the laser paraxial focal plane. By symmetry, $E^2,$ $Q_z$ and $Q_{\rho}$ are independent of $\phi.$ 

The electric energy density  depicted in the first line of Fig.~\ref{Mosaic}(a) spreads out radially and in the region below the laser paraxial focal plane as a result of the increase of the spherical aberration 
phase (\ref{phi_SA}). The density plots of the optical force components allow us to identify the roots of $Q_{\rho}=0$ and $Q_z=0$ leading to stable trapping. They are indicated as green and blue lines, respectively. 
The trapping configurations are obtained as the intersections between the two lines. 

In the absence of spherical aberration ($L=0$), the green and blue lines intersect at an axial position as well as off-axis, in line with the result of Fig.~\ref{phase_diag} since the parameters 
$(\ell=8,a=1.5\,\mu{\rm m})$ lie within the bistable colored region. In order to understand why the microsphere stays on the axis in this case, we plot in Fig.~\ref{Mosaic}(b) the optical potential 
$U(\rho)\equiv -\int_0^{\rho} F_{\rho}(\rho,\bar{z}(\rho)) d\rho,$ where $\bar{z}(\rho)$ is the axial coordinate leading to a vanishing axial force at  $\rho$: $Q_z(\rho,\bar{z}(\rho))=0$ (the function $\bar{z}(\rho)$ 
corresponds to the blue lines in Fig.~\ref{Mosaic}(a)). In order to calculate the radial force component $F_{\rho},$ we consider the expressions for $Q_{\rho}$ given in Appendix~\ref{sec:explicit} and determine the 
power at the sample $P$ from the period of rotation (see Appendix \ref{sec:Faxen} for details). The potential $U(\rho)$ corresponds to the conservative component of the optical force field~\cite{Roichman2008,Huang2022} 
and provides a qualitative indication of the different trapping regimes. Indeed, the leftmost plot in  Fig.~\ref{Mosaic}(b) indicates that the well at $\rho=0$ is much deeper than the one corresponding to off-axis orbital 
motion, which explains the experimental observation of stable axial trapping. As $L$ increases, the off-axis well gets deeper, leading to the bistable trapping near $L=1.75\,\mu{\rm m}.$ When compared to the experimental 
energy distribution shown in Fig.~\ref{FigExpData}(f), theory overestimates the difference between the local minima by a factor $\sim 2.$ The axial equilibrium position eventually vanishes as the focal spot continues to 
spread out by increasing $L,$ and then the microsphere stays on its off-axis orbital motion at $L=3.5\,\mu{\rm m}$ (rightmost column in Fig.~\ref{Mosaic}).

The difference between the microsphere image patterns for on-axis  and off-axis orbital motion illustrated by Figs.~\ref{FigExpData}(a)-(d) is in qualitative agreement with the difference between the equilibrium values 
$z_{\rm axis}$ and $z_{\rm orb},$ which are obtained from the intersections between the green and blue lines shown in the density plot for $L=1.75\,\mu{\rm m}$ in Fig.~\ref{Mosaic}(a). Fig.~\ref{FigExpData}(f) shows that 
the distribution $p(x,y)$ near the axis is elongated along the direction bisecting the first and third quadrants of the $xy$ plane, while the orbit is extended along the orthogonal direction. Those properties are consistent 
with astigmatism of the trapping beam~\cite{Roichman2006,Dutra2014}, with the plane of least confusion located in between the planes of the orbital motion and of the on-axis equilibrium. Fig.~\ref{FigExpData}(f) indicates 
that astigmatism plays an important role in the transitions between axial and orbital states. Indeed, the position distribution shows that the particle preferably hops to and from orbital microstates closer to the axial state. 
Such bias is not captured by our stigmatic model, which requires unbiased thermal fluctuations to break the rotational symmetry when hopping from the axial state to a given microstate along the circular orbit. In addition, 
we surmise that astigmatism decreases the potential barrier separating the two minima shown in the potential for $L=1.75\,\mu{\rm m}$ shown in Fig.~\ref{Mosaic}(b), thus facilitating thermally-activated hops along both ways as 
observed in the experiment.

The  theoretical results for the orbital radii are organized in Table I, with the errors arising from the uncertainty of $L.$ Although our model does not take astigmatism into account, we still find good agreement with the 
experimental data for $\ell=5$ and $\ell=8.$ Indeed, it is generally expected that particles with radii $a>\lambda_0$ average out the imperfections arising from astigmatism~\cite{Dutra2014}. However, $\lambda_0$ is replaced by 
the radius of the annular spot $\tilde{r}_{\ell}>\lambda_0$ as the characteristic size of the diffraction-limited focal spot when employing vortex beams. Thus, we attribute the agreement mostly to time averaging the radial distance 
over several periods of revolution, which effectively averages out the elongation of the orbit shown in Fig.~\ref{FigExpData}(f). 

For $\ell=4,$ our theoretical model predicts axial trapping only, regardless of the amount of spherical aberration. In other words, in this case the prediction of stable axial trapping with an aplanatic focused beam (note that 
$a=1.0\,\mu{\rm}{\rm m} > R_{\rm off}(\ell=4)$ as indicated in Fig.~\ref{phase_diag}) is not modified by the introduction of spherical aberration.

\section{Conclusion}
\label{sec:conclusion}

In conclusion, we have demonstrated bistable optical trapping by employing a vortex beam at the objective back aperture. The non-equilibrium steady-state corresponding to orbital motion driven by the optical torque co-exists with 
stable axial trapping. The corresponding bistable optical potential has the form of a Mexican hat with an additional minimum at its center. To achieve such bistable trapping, the microsphere diameter should be comparable with the 
diameter of the laser focal spot, which has an annular shape in the case of circular polarization.  

Our experimental results are compared with an extension of MDSA theory of optical tweezers considering a circularly-polarized vortex beam at the objective back aperture. The ideal case of a an aplanatic focused beam provides a 
useful guide in the search for bistable behavior. In particular, it shows that the range of  particle sizes yielding bistability becomes wider as the topological charge increases. However, spherical aberration is essential for a full 
description of our experiment employing an oil-immersion objective. More importantly, spherical aberration allows us to to tailor different trapping regimes. 
Indeed, since the focal height can be precisely controlled with the help of a piezoelectric nanopositioning system, our system allows us to manipulate the transition from a single state,  either on the optical axis or in a well-defined 
orbit, to a metastable state. The cyclic hops between the two mesostates, each with a significantly distinct set of microstates, open the way for investigating the energetics of cyclic symmetry breaking and restoration \cite{Roldan2014}.
The possibility of employing a time-dependent focal height could also find applications in shortcuts to equilibration~\cite{Martinez2016,Raynal2023,Pires2023} connecting mesostates with different symmetries.

\section*{Acknowledgments}

We are grateful to C\'assia S. Nascimento, Cyriaque Genet, Fran\c cois Impens, Gert-Ludwig Ingold, R\'emi Goerlich and Tanja Schoger for fruitful discussions. A.~L.~F. expresses his deepest gratitude to the Light-matter Interactions 
and Nanostructures group at Institut de Science et d’Ingénierie Supramoléculaire (ISIS, Strasbourg, France) for their hospitality during his stay as a visiting student. P.~A.~M.~N. acknowledges funding from the Brazilian agencies Conselho 
Nacional de Desenvolvimento Cient\'{\i}fico e Tecnol\'ogico (CNPq--Brazil), Coordenaç\~ao de Aperfeiçamento de Pessoal de N\'{\i}vel Superior (CAPES--Brazil),  Instituto Nacional de Ci\^encia e Tecnologia Fluidos Complexos  (INCT-FCx), and 
the Research Foundations of the States of Rio de Janeiro (FAPERJ) and S\~ao Paulo (FAPESP).

\appendix

\section{Objective filling  }
\label{sec:beam waist}
Objective filling conditions are particularly important when using a vortex beam as we need to optimize the energy density gradient (by reducing the size of the focal spot) while keeping most of the annular section of the incoming beam 
inside the objective entrance port. Filling is controlled by the ratio $r_\ell/R_p$ between the radius of the vortex beam 
\begin{equation}
    r_\ell = w_0 \sqrt{\frac{|\ell|}{2}}
    \label{int_max_lg}
\end{equation}
and the radius of the objective entrance port $R_p.$

\begin{figure}[ht]
    \includegraphics[scale=.53]{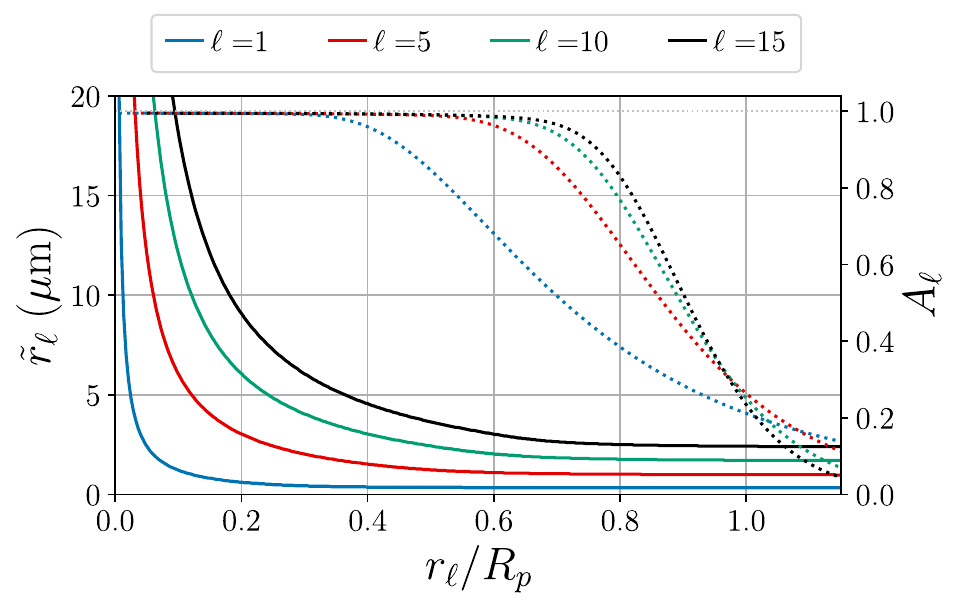}
    \caption{Radius $\tilde{r}_\ell$ of the focal annular spot (solid, left axis)  and 
    filling factor $A_\ell$ representing the fraction of total power transmitted to the sample chamber (dot, right axis) as functions of the radius $r_\ell$ of the paraxial vortex beam (in units of the objective entrance radius $R_p=2.8\,{\rm mm}$). 
    The topological charges are $\ell=1,5,10,15$. We consider an oil-immersion objective lens with numerical aperture $\rm{NA}=1.4$. }
    \label{tilderell_Aell_rell}
\end{figure}

In Fig.~\ref{tilderell_Aell_rell}, we plot the radius $\tilde{r}_\ell$ of the focal annular spot (solid) and the filling factor $A_{\ell}$ (dot), as given by Eq.~(\ref{Al_SA}), as functions of $r_\ell/R_p$ for different values of $\ell.$ 
We consider the oil-immersion objective used in the experiment (see Sec.~\ref{sec:experiment} for details). 

As expected, the fraction of the total power that enters the objective entrance port decreases as the radius of the vortex beam $r_{\ell}$ increases. The focal spot radius $\tilde{r}_\ell$ also decreases with $r_\ell/R_p,$ reaching 
diffraction-limited values (which scale linearly with $\ell$~\cite{Curtis2003,Monteiro2009}) at $r_\ell/R_p\stackrel{>}{\scriptscriptstyle\sim} 0.8.$ In order to simulate a diffraction-limited spot with the minimal power loss, we take 
$r_\ell/R_p = 0.8$ for the aberration-free calculations presented in Sec.~\ref{sec:Trapping regimes}. The values of  $r_\ell/R_p$ corresponding to the experiment, shown in Table \ref{table_data_i}, are employed for the calculation of the 
MDSA results which are compared to the experimental data.

In order to stay  close to the optimal filling condition, we produce vortex beams with increasingly smaller values of the waist $w_0$ as $\ell$ increases, as indicated in Table~\ref{table_data_i}. Such condition also allows  the vortex 
beam to develop a finite curvature as it propagates towards the objective back aperture. The resulting curvature of field shifts the whole laser focal spot upwards without changing the amount of spherical aberration~\cite{Born2019}, thus 
making room for trapping above the glass slide in spite of the small values of the objective focal height $L$ employed in the experiment. 

\begin{table}[ht]
\begin{tabularx}{0.48\textwidth}{|| >{\centering\hsize=.25\hsize\linewidth=\hsize}X | >{\centering\hsize=.75\hsize}X | >{\centering\arraybackslash}X | >{\centering\arraybackslash}X | >{\centering\arraybackslash}X | >{\centering\arraybackslash}X ||} 
 \hline
 $\ell$ & $a\,(\mu\rm{m})$   & $w_0\,(\rm{mm})$ & $r_\ell/R_p$
 \\  [0.2ex]
 \hline\hline
 $4$ & $1.0$   & $1.78\pm0.02$  & $0.90\pm0.01$\\
 \hline
 $5$ & $1.0$   & $1.83\pm0.02$  & $1.03\pm0.01$\\ 
 \hline
 $8$ & $1.5$   & $0.95\pm0.01$  & $0.68\pm0.01$ \\
 \hline
 $11$ & $2.25$ &  $0.89\pm0.01$ & $0.74\pm0.01$ \\ [.1ex] 
 \hline
\end{tabularx}
\caption{Vortex beam waist $w_0$ and ratio between the radii of the beam (at the objective entrance port) and of the objective entrance port $R_p=(2.80 \pm 0.05)\, {\rm mm}$. }
\label{table_data_i}
\end{table}

\section{Partial-wave (multipole) series for the optical force cylindrical components}
\label{sec:explicit}

The radial and azimuthal components of the extinction contribution to the optical force are given by
\begin{widetext}
\begin{align} 
        \begin{split}
        Q_{e\rho} = \frac{(2\gamma^2)^{|\ell|+1}}{ |\ell|! \, A_{\ell} \, N_a}\Im\sum_{jm}\Big[(2j+1)(a_j + b_j)
        G_{j,m}\big(G^{(-)}_{j,m+1}-G^{(+)}_{j,m-1}\big)^*\Big]
        \label{Qerho}
        \end{split}
    \end{align}
\begin{align} 
        \begin{split}
        Q_{e\phi} =& \frac{-(2\gamma^2)^{|\ell|+1}}{ |\ell|! \, A_{\ell} \, N_a}\Re\sum_{jm}\Big[(2j+1)(a_j + b_j) G_{j,m}\big(G^{(-)}_{j,m+1}+G^{(+)}_{j,m-1}\big)^*\Big].
        \label{Qephi}
        \end{split}
    \end{align}
Here, we have defined the additional coefficients

    \begin{equation}
        \begin{split}
        G^\pm_{jm} = \int^{\theta_0}_0 d\theta\, \sin\theta_1(\sin\theta)^{1+|\ell|}\sqrt{\cos\theta}\, e^{-\gamma^2 \sin^2\theta} e^{i \Phi(z, \theta)} T_\perp(\theta_1)d^j_{m \pm 1,1}(\theta_1)
        J_{m-1-\ell}(k\rho\sin\theta_1).
        \label{Gpm}
        \end{split}
    \end{equation}

Finally, the scattering contribution to the optical force is written in terms of cylindrical components as follows:

    \begin{align} 
        \begin{split}
        Q_{s\rho} =\! \frac{2(2\gamma^2)^{|\ell|+1}}{ |\ell|! \, A_{\ell} \, N_a} 
        \sum_{jm}\!\Bigg[\!
        \frac{\sqrt{j(j+2)(j+m+1)(j+m+2)}}{(j+1)}
        &\Im\Big( (a_j a_{j+1}^*+b_j b_{j+1}^*) 
        \big(G_{j,m}G^*_{j+1,m+1} + G_{j,-m}G^*_{j+1,-m-1}\big)\Big)\\
        -2\frac{(2j+1)}{j(j+1)}\sqrt{(j-m)(j+m+1)} &\Re(a_jb_j^*)\Im\big(G_{j,m}G^*_{j,m+1}\big)\Bigg]
        \label{Qsrho}
        \end{split}
    \end{align}

    \begin{align} 
        \begin{split}
        Q_{s\phi} =\! \frac{-2(2\gamma^2)^{|\ell|+1}}{ |\ell|! \, A_{\ell} \, N_a}\!\sum_{jm}\!\Bigg[\!
        \frac{\sqrt{j(j+2)(j+m+1)(j+m+2)}}{(j+1)} 
        &\Re\!\Big((a_j a_{j+1}^*\! + b_j b_{j+1}^*) \big(G_{j,m}G^*_{j+1,m+1}\! - G_{j,-m}G^*_{j+1,-m-1} \big) \Big)\\
        -2\frac{(2j+1)}{j(j+1)}\sqrt{(j-m)(j+m+1)} &\Re(a_jb_j^*)\Re\big(G_{j,m}G^*_{j,m+1}\big)\Bigg]
        \label{Qsphi}
        \end{split}
    \end{align}

    \begin{align} 
        \begin{split}
        Q_{sz} = \frac{-4(2\gamma^2)^{|\ell|+1}}{ |\ell|! \, A_{\ell} \, N_a}\sum_{jm}\Bigg[
        &\frac{\sqrt{j(j+2)(j-m+1)(j+m+1)}}{(j+1)}
        \Re\Big((a_j a_{j+1}^* + b_j b_{j+1}^*) \big(G_{j,m}G^*_{j+1,m}\big)\\
        &+\frac{(2j+1)}{j(j+1)}\,m(a_j b_j^*)\big(G_{j,m}G^*_{j,m}\big)\Big)
        \Bigg].
        \end{split}
        \label{Qsz}
    \end{align}

\end{widetext}

\section{Laser power in the sample region }
\label{sec:Faxen}

Due to the non-uniform transmittance~\cite{Viana2006} of our high-NA objective, we were not able to estimate the laser power $P$ delivered to the sample from the power at the objective 
entrance port. Instead, we determine $P$ from the period of rotation $T_{\rm exp}$ by taking the Stokes friction force along the orbital motion to match the azimuthal component of the 
optical force: $\beta\,v_\phi=n_1P Q_\phi/ c,$ where $v_\phi=2\pi \rho_{\rm exp}/T_{\rm exp}$ is the velocity along the orbit. The MDSA multipole expansion for $Q_\phi$ is given by 
Eqs.~(\ref{Qephi}) and (\ref{Qsphi}) of Appendix \ref{sec:explicit}. We take Fax\'en's correction arising from the glass slide when evaluating the drag coefficient $\beta$~\cite{Feitosa1991,Schaffer2007}:
\begin{equation}
    \beta = \frac{6\pi\eta a}{1-\frac{9}{16}\left(\frac{a}{h}\right)+\frac{1}{8}\left(\frac{a}{h}\right)^3-\frac{45}{256}\left(\frac{a}{h}\right)^4-\frac{1}{16}\left(\frac{a}{h}\right)^5},
    \label{Faxen}
\end{equation}
where $h$ is the height of the microsphere center with respect to the glass slide and $\eta=(0.91\pm0.02)\,\rm{mPa}\cdot\rm{s}$ is the  viscosity of water at room temperature ($298\,$K).

In order to estimate $h,$ we start each experimental run with a reference configuration such that the trapped microsphere is barely touching the glass slide. We then displace the sample downwards by a 
distance $d=2\,\mu{\rm m}$. Neglecting the variation of the axial trapping position with respect to the laser focal plane, we have  $h=a+N_ad.$ For $\ell=8$ and $a=1.5\,\mu{\rm m},$ we find  
$\beta=(30\pm1)\,\mu{\rm g}/{\rm s}$ from Eq.~(\ref{Faxen}). The azimuthal optical force is given by $Q_\phi=0.0212$ and then the resulting power is $P=3.8\,{\rm mW}.$


\begin{thebibliography}{56}%
\makeatletter
\providecommand \@ifxundefined [1]{%
 \@ifx{#1\undefined}
}%
\providecommand \@ifnum [1]{%
 \ifnum #1\expandafter \@firstoftwo
 \else \expandafter \@secondoftwo
 \fi
}%
\providecommand \@ifx [1]{%
 \ifx #1\expandafter \@firstoftwo
 \else \expandafter \@secondoftwo
 \fi
}%
\providecommand \natexlab [1]{#1}%
\providecommand \enquote  [1]{``#1''}%
\providecommand \bibnamefont  [1]{#1}%
\providecommand \bibfnamefont [1]{#1}%
\providecommand \citenamefont [1]{#1}%
\providecommand \href@noop [0]{\@secondoftwo}%
\providecommand \href [0]{\begingroup \@sanitize@url \@href}%
\providecommand \@href[1]{\@@startlink{#1}\@@href}%
\providecommand \@@href[1]{\endgroup#1\@@endlink}%
\providecommand \@sanitize@url [0]{\catcode `\\12\catcode `\$12\catcode
  `\&12\catcode `\#12\catcode `\^12\catcode `\_12\catcode `\%12\relax}%
\providecommand \@@startlink[1]{}%
\providecommand \@@endlink[0]{}%
\providecommand \url  [0]{\begingroup\@sanitize@url \@url }%
\providecommand \@url [1]{\endgroup\@href {#1}{\urlprefix }}%
\providecommand \urlprefix  [0]{URL }%
\providecommand \Eprint [0]{\href }%
\providecommand \doibase [0]{https://doi.org/}%
\providecommand \selectlanguage [0]{\@gobble}%
\providecommand \bibinfo  [0]{\@secondoftwo}%
\providecommand \bibfield  [0]{\@secondoftwo}%
\providecommand \translation [1]{[#1]}%
\providecommand \BibitemOpen [0]{}%
\providecommand \bibitemStop [0]{}%
\providecommand \bibitemNoStop [0]{.\EOS\space}%
\providecommand \EOS [0]{\spacefactor3000\relax}%
\providecommand \BibitemShut  [1]{\csname bibitem#1\endcsname}%
\let\auto@bib@innerbib\@empty
\bibitem [{\citenamefont {Ashkin}\ \emph {et~al.}(1986)\citenamefont {Ashkin},
  \citenamefont {Dziedzic}, \citenamefont {Bjorkholm},\ and\ \citenamefont
  {Chu}}]{Ashkin1986}%
  \BibitemOpen
  \bibfield  {author} {\bibinfo {author} {\bibfnamefont {A.}~\bibnamefont
  {Ashkin}}, \bibinfo {author} {\bibfnamefont {J.~M.}\ \bibnamefont
  {Dziedzic}}, \bibinfo {author} {\bibfnamefont {J.~E.}\ \bibnamefont
  {Bjorkholm}},\ and\ \bibinfo {author} {\bibfnamefont {S.}~\bibnamefont
  {Chu}},\ }\bibfield  {title} {\bibinfo {title} {{Observation of a single-beam
  gradient force optical trap for dielectric particles}},\ }\href@noop {}
  {\bibfield  {journal} {\bibinfo  {journal} {Optics Letters}\ }\textbf
  {\bibinfo {volume} {11}},\ \bibinfo {pages} {288} (\bibinfo {year}
  {1986})}\BibitemShut {NoStop}%
\bibitem [{\citenamefont {Ashkin}(2006)}]{Ashkin2006}%
  \BibitemOpen
  \bibfield  {author} {\bibinfo {author} {\bibfnamefont {A.}~\bibnamefont
  {Ashkin}},\ }\href@noop {} {\emph {\bibinfo {title} {{Optical Trapping and
  Manipulation of Neutral Particles Using Lasers: a Reprint Volume with
  Commentaries}}}}\ (\bibinfo  {publisher} {World Scientific},\ \bibinfo {year}
  {2006})\BibitemShut {NoStop}%
\bibitem [{\citenamefont {Gennerich}(2016)}]{Gennerich2016}%
  \BibitemOpen
  \bibfield  {author} {\bibinfo {author} {\bibfnamefont {A.}~\bibnamefont
  {Gennerich}},\ }\href {https://books.google.com.br/books?id=T-mynQAACAAJ}
  {\emph {\bibinfo {title} {Optical Tweezers: Methods and Protocols}}},\
  Methods in Molecular Biology\ (\bibinfo  {publisher} {Springer New York},\
  \bibinfo {year} {2016})\BibitemShut {NoStop}%
\bibitem [{\citenamefont {Forbes}\ \emph {et~al.}(2021)\citenamefont {Forbes},
  \citenamefont {de~Oliveira},\ and\ \citenamefont {Dennis}}]{Forbes2021}%
  \BibitemOpen
  \bibfield  {author} {\bibinfo {author} {\bibfnamefont {A.}~\bibnamefont
  {Forbes}}, \bibinfo {author} {\bibfnamefont {M.}~\bibnamefont
  {de~Oliveira}},\ and\ \bibinfo {author} {\bibfnamefont {M.~R.}\ \bibnamefont
  {Dennis}},\ }\bibfield  {title} {\bibinfo {title} {Structured light},\
  }\href@noop {} {\bibfield  {journal} {\bibinfo  {journal} {Nature Photonics}\
  }\textbf {\bibinfo {volume} {15}},\ \bibinfo {pages} {253} (\bibinfo {year}
  {2021})}\BibitemShut {NoStop}%
\bibitem [{\citenamefont {Rubinsztein-Dunlop}\ \emph
  {et~al.}(2016)\citenamefont {Rubinsztein-Dunlop}, \citenamefont {Forbes},
  \citenamefont {Berry}, \citenamefont {Dennis}, \citenamefont {Andrews},
  \citenamefont {Mansuripur}, \citenamefont {Denz}, \citenamefont {Alpmann},
  \citenamefont {Banzer}, \citenamefont {Bauer}, \citenamefont {Karimi},
  \citenamefont {Marrucci}, \citenamefont {Padgett}, \citenamefont
  {Ritsch-Marte}, \citenamefont {Litchinitser}, \citenamefont {Bigelow},
  \citenamefont {Rosales-Guzmán}, \citenamefont {Belmonte}, \citenamefont
  {Torres}, \citenamefont {Neely}, \citenamefont {Baker}, \citenamefont
  {Gordon}, \citenamefont {Stilgoe}, \citenamefont {Romero}, \citenamefont
  {White}, \citenamefont {Fickler}, \citenamefont {Willner}, \citenamefont
  {Xie}, \citenamefont {McMorran},\ and\ \citenamefont
  {Weiner}}]{Rubinsztein-Dunlop2017}%
  \BibitemOpen
  \bibfield  {author} {\bibinfo {author} {\bibfnamefont {H.}~\bibnamefont
  {Rubinsztein-Dunlop}}, \bibinfo {author} {\bibfnamefont {A.}~\bibnamefont
  {Forbes}}, \bibinfo {author} {\bibfnamefont {M.~V.}\ \bibnamefont {Berry}},
  \bibinfo {author} {\bibfnamefont {M.~R.}\ \bibnamefont {Dennis}}, \bibinfo
  {author} {\bibfnamefont {D.~L.}\ \bibnamefont {Andrews}}, \bibinfo {author}
  {\bibfnamefont {M.}~\bibnamefont {Mansuripur}}, \bibinfo {author}
  {\bibfnamefont {C.}~\bibnamefont {Denz}}, \bibinfo {author} {\bibfnamefont
  {C.}~\bibnamefont {Alpmann}}, \bibinfo {author} {\bibfnamefont
  {P.}~\bibnamefont {Banzer}}, \bibinfo {author} {\bibfnamefont
  {T.}~\bibnamefont {Bauer}}, \bibinfo {author} {\bibfnamefont
  {E.}~\bibnamefont {Karimi}}, \bibinfo {author} {\bibfnamefont
  {L.}~\bibnamefont {Marrucci}}, \bibinfo {author} {\bibfnamefont
  {M.}~\bibnamefont {Padgett}}, \bibinfo {author} {\bibfnamefont
  {M.}~\bibnamefont {Ritsch-Marte}}, \bibinfo {author} {\bibfnamefont {N.~M.}\
  \bibnamefont {Litchinitser}}, \bibinfo {author} {\bibfnamefont {N.~P.}\
  \bibnamefont {Bigelow}}, \bibinfo {author} {\bibfnamefont {C.}~\bibnamefont
  {Rosales-Guzmán}}, \bibinfo {author} {\bibfnamefont {A.}~\bibnamefont
  {Belmonte}}, \bibinfo {author} {\bibfnamefont {J.~P.}\ \bibnamefont
  {Torres}}, \bibinfo {author} {\bibfnamefont {T.~W.}\ \bibnamefont {Neely}},
  \bibinfo {author} {\bibfnamefont {M.}~\bibnamefont {Baker}}, \bibinfo
  {author} {\bibfnamefont {R.}~\bibnamefont {Gordon}}, \bibinfo {author}
  {\bibfnamefont {A.~B.}\ \bibnamefont {Stilgoe}}, \bibinfo {author}
  {\bibfnamefont {J.}~\bibnamefont {Romero}}, \bibinfo {author} {\bibfnamefont
  {A.~G.}\ \bibnamefont {White}}, \bibinfo {author} {\bibfnamefont
  {R.}~\bibnamefont {Fickler}}, \bibinfo {author} {\bibfnamefont {A.~E.}\
  \bibnamefont {Willner}}, \bibinfo {author} {\bibfnamefont {G.}~\bibnamefont
  {Xie}}, \bibinfo {author} {\bibfnamefont {B.}~\bibnamefont {McMorran}},\ and\
  \bibinfo {author} {\bibfnamefont {A.~M.}\ \bibnamefont {Weiner}},\ }\bibfield
   {title} {\bibinfo {title} {Roadmap on structured light},\ }\href
  {https://doi.org/10.1088/2040-8978/19/1/013001} {\bibfield  {journal}
  {\bibinfo  {journal} {Journal of Optics}\ }\textbf {\bibinfo {volume} {19}},\
  \bibinfo {pages} {013001} (\bibinfo {year} {2016})}\BibitemShut {NoStop}%
\bibitem [{\citenamefont {Shen}\ \emph {et~al.}(2019)\citenamefont {Shen},
  \citenamefont {Wang}, \citenamefont {Xie}, \citenamefont {Min}, \citenamefont
  {Fu}, \citenamefont {Liu}, \citenamefont {Gong},\ and\ \citenamefont
  {Yuan}}]{Shen2019}%
  \BibitemOpen
  \bibfield  {author} {\bibinfo {author} {\bibfnamefont {Y.}~\bibnamefont
  {Shen}}, \bibinfo {author} {\bibfnamefont {X.}~\bibnamefont {Wang}}, \bibinfo
  {author} {\bibfnamefont {Z.}~\bibnamefont {Xie}}, \bibinfo {author}
  {\bibfnamefont {C.}~\bibnamefont {Min}}, \bibinfo {author} {\bibfnamefont
  {X.}~\bibnamefont {Fu}}, \bibinfo {author} {\bibfnamefont {Q.}~\bibnamefont
  {Liu}}, \bibinfo {author} {\bibfnamefont {M.}~\bibnamefont {Gong}},\ and\
  \bibinfo {author} {\bibfnamefont {X.}~\bibnamefont {Yuan}},\ }\bibfield
  {title} {\bibinfo {title} {Optical vortices 30 years on: Oam manipulation
  from topological charge to multiple singularities},\ }\href@noop {}
  {\bibfield  {journal} {\bibinfo  {journal} {Light: Science \& Applications}\
  }\textbf {\bibinfo {volume} {8}},\ \bibinfo {pages} {90} (\bibinfo {year}
  {2019})}\BibitemShut {NoStop}%
\bibitem [{\citenamefont {Yang}\ \emph {et~al.}(2021)\citenamefont {Yang},
  \citenamefont {Ren}, \citenamefont {Chen}, \citenamefont {Arita},\ and\
  \citenamefont {Rosales-Guzm{\'{a}}n}}]{Yang2021}%
  \BibitemOpen
  \bibfield  {author} {\bibinfo {author} {\bibfnamefont {Y.}~\bibnamefont
  {Yang}}, \bibinfo {author} {\bibfnamefont {Y.-X.}\ \bibnamefont {Ren}},
  \bibinfo {author} {\bibfnamefont {M.}~\bibnamefont {Chen}}, \bibinfo {author}
  {\bibfnamefont {Y.}~\bibnamefont {Arita}},\ and\ \bibinfo {author}
  {\bibfnamefont {C.}~\bibnamefont {Rosales-Guzm{\'{a}}n}},\ }\bibfield
  {title} {\bibinfo {title} {Optical trapping with structured light: a
  review},\ }\bibfield  {journal} {\bibinfo  {journal} {Advanced Photonics}\
  }\textbf {\bibinfo {volume} {3}},\ \href
  {https://doi.org/10.1117/1.ap.3.3.034001} {10.1117/1.ap.3.3.034001} (\bibinfo
  {year} {2021})\BibitemShut {NoStop}%
\bibitem [{\citenamefont {Li}\ \emph {et~al.}(2021{\natexlab{a}})\citenamefont
  {Li}, \citenamefont {Yan}, \citenamefont {Zhang}, \citenamefont {Chen},\ and\
  \citenamefont {Yao}}]{Li2021b}%
  \BibitemOpen
  \bibfield  {author} {\bibinfo {author} {\bibfnamefont {M.}~\bibnamefont
  {Li}}, \bibinfo {author} {\bibfnamefont {S.}~\bibnamefont {Yan}}, \bibinfo
  {author} {\bibfnamefont {Y.}~\bibnamefont {Zhang}}, \bibinfo {author}
  {\bibfnamefont {X.}~\bibnamefont {Chen}},\ and\ \bibinfo {author}
  {\bibfnamefont {B.}~\bibnamefont {Yao}},\ }\bibfield  {title} {\bibinfo
  {title} {Optical separation and discrimination of chiral particles by vector
  beams with orbital angular momentum},\ }\href
  {https://doi.org/10.1039/d1na00530h} {\bibfield  {journal} {\bibinfo
  {journal} {Nanoscale Advances}\ }\textbf {\bibinfo {volume} {3}},\ \bibinfo
  {pages} {6897} (\bibinfo {year} {2021}{\natexlab{a}})}\BibitemShut {NoStop}%
\bibitem [{\citenamefont {Bobkova}\ \emph {et~al.}(2021)\citenamefont
  {Bobkova}, \citenamefont {Stegemann}, \citenamefont {Droop}, \citenamefont
  {Otte},\ and\ \citenamefont {Denz}}]{Bobkova2021}%
  \BibitemOpen
  \bibfield  {author} {\bibinfo {author} {\bibfnamefont {V.}~\bibnamefont
  {Bobkova}}, \bibinfo {author} {\bibfnamefont {J.}~\bibnamefont {Stegemann}},
  \bibinfo {author} {\bibfnamefont {R.}~\bibnamefont {Droop}}, \bibinfo
  {author} {\bibfnamefont {E.}~\bibnamefont {Otte}},\ and\ \bibinfo {author}
  {\bibfnamefont {C.}~\bibnamefont {Denz}},\ }\bibfield  {title} {\bibinfo
  {title} {Optical grinder: sorting of trapped particles by orbital angular
  momentum},\ }\href {https://doi.org/10.1364/OE.419876} {\bibfield  {journal}
  {\bibinfo  {journal} {Opt. Express}\ }\textbf {\bibinfo {volume} {29}},\
  \bibinfo {pages} {12967} (\bibinfo {year} {2021})}\BibitemShut {NoStop}%
\bibitem [{\citenamefont {Spreeuw}(2022)}]{Spreeuw2022}%
  \BibitemOpen
  \bibfield  {author} {\bibinfo {author} {\bibfnamefont {R.~J.~C.}\
  \bibnamefont {Spreeuw}},\ }\bibfield  {title} {\bibinfo {title} {Spiraling
  light: from donut modes to a magnus effect analogy},\ }\href
  {https://doi.org/doi:10.1515/nanoph-2021-0458} {\bibfield  {journal}
  {\bibinfo  {journal} {Nanophotonics}\ }\textbf {\bibinfo {volume} {11}},\
  \bibinfo {pages} {633} (\bibinfo {year} {2022})}\BibitemShut {NoStop}%
\bibitem [{\citenamefont {Allen}\ \emph {et~al.}(1992)\citenamefont {Allen},
  \citenamefont {Beijersbergen}, \citenamefont {Spreeuw},\ and\ \citenamefont
  {Woerdman}}]{Allen1992}%
  \BibitemOpen
  \bibfield  {author} {\bibinfo {author} {\bibfnamefont {L.}~\bibnamefont
  {Allen}}, \bibinfo {author} {\bibfnamefont {M.~W.}\ \bibnamefont
  {Beijersbergen}}, \bibinfo {author} {\bibfnamefont {R.~J.~C.}\ \bibnamefont
  {Spreeuw}},\ and\ \bibinfo {author} {\bibfnamefont {J.~P.}\ \bibnamefont
  {Woerdman}},\ }\bibfield  {title} {\bibinfo {title} {Orbital angular momentum
  of light and the transformation of laguerre-gaussian laser modes},\ }\href
  {https://doi.org/10.1103/PhysRevA.45.8185} {\bibfield  {journal} {\bibinfo
  {journal} {Phys. Rev. A}\ }\textbf {\bibinfo {volume} {45}},\ \bibinfo
  {pages} {8185} (\bibinfo {year} {1992})}\BibitemShut {NoStop}%
\bibitem [{\citenamefont {He}\ \emph {et~al.}(1995)\citenamefont {He},
  \citenamefont {Friese}, \citenamefont {Heckenberg},\ and\ \citenamefont
  {Rubinsztein-Dunlop}}]{He1995}%
  \BibitemOpen
  \bibfield  {author} {\bibinfo {author} {\bibfnamefont {H.}~\bibnamefont
  {He}}, \bibinfo {author} {\bibfnamefont {M.~E.~J.}\ \bibnamefont {Friese}},
  \bibinfo {author} {\bibfnamefont {N.~R.}\ \bibnamefont {Heckenberg}},\ and\
  \bibinfo {author} {\bibfnamefont {H.}~\bibnamefont {Rubinsztein-Dunlop}},\
  }\bibfield  {title} {\bibinfo {title} {Direct observation of transfer of
  angular momentum to absorptive particles from a laser beam with a phase
  singularity},\ }\href {https://doi.org/10.1103/PhysRevLett.75.826} {\bibfield
   {journal} {\bibinfo  {journal} {Phys. Rev. Lett.}\ }\textbf {\bibinfo
  {volume} {75}},\ \bibinfo {pages} {826} (\bibinfo {year} {1995})}\BibitemShut
  {NoStop}%
\bibitem [{\citenamefont {Simpson}\ \emph {et~al.}(1997)\citenamefont
  {Simpson}, \citenamefont {Dholakia}, \citenamefont {Allen},\ and\
  \citenamefont {Padgett}}]{Simpson1997}%
  \BibitemOpen
  \bibfield  {author} {\bibinfo {author} {\bibfnamefont {N.~B.}\ \bibnamefont
  {Simpson}}, \bibinfo {author} {\bibfnamefont {K.}~\bibnamefont {Dholakia}},
  \bibinfo {author} {\bibfnamefont {L.}~\bibnamefont {Allen}},\ and\ \bibinfo
  {author} {\bibfnamefont {M.~J.}\ \bibnamefont {Padgett}},\ }\bibfield
  {title} {\bibinfo {title} {Mechanical equivalence of spin and orbital angular
  momentum of light: an optical spanner},\ }\href
  {https://doi.org/10.1364/OL.22.000052} {\bibfield  {journal} {\bibinfo
  {journal} {Opt. Lett.}\ }\textbf {\bibinfo {volume} {22}},\ \bibinfo {pages}
  {52} (\bibinfo {year} {1997})}\BibitemShut {NoStop}%
\bibitem [{\citenamefont {Bliokh}\ \emph {et~al.}(2011)\citenamefont {Bliokh},
  \citenamefont {Ostrovskaya}, \citenamefont {Alonso}, \citenamefont
  {Rodr{\'{i}}guez-Herrera}, \citenamefont {Lara},\ and\ \citenamefont
  {Dainty}}]{Bliokh2011}%
  \BibitemOpen
  \bibfield  {author} {\bibinfo {author} {\bibfnamefont {K.~Y.}\ \bibnamefont
  {Bliokh}}, \bibinfo {author} {\bibfnamefont {E.~A.}\ \bibnamefont
  {Ostrovskaya}}, \bibinfo {author} {\bibfnamefont {M.~A.}\ \bibnamefont
  {Alonso}}, \bibinfo {author} {\bibfnamefont {O.~G.}\ \bibnamefont
  {Rodr{\'{i}}guez-Herrera}}, \bibinfo {author} {\bibfnamefont
  {D.}~\bibnamefont {Lara}},\ and\ \bibinfo {author} {\bibfnamefont
  {C.}~\bibnamefont {Dainty}},\ }\bibfield  {title} {\bibinfo {title}
  {{Spin-to-orbital angular momentum conversion in focusing, scattering, and
  imaging systems}},\ }\href@noop {} {\bibfield  {journal} {\bibinfo  {journal}
  {Optics Express}\ }\textbf {\bibinfo {volume} {19}},\ \bibinfo {pages}
  {26132} (\bibinfo {year} {2011})}\BibitemShut {NoStop}%
\bibitem [{\citenamefont {Bliokh}\ \emph {et~al.}(2015)\citenamefont {Bliokh},
  \citenamefont {Rodr{\'{i}}guez-Fortu{\~{n}}o}, \citenamefont {Nori},\ and\
  \citenamefont {Zayats}}]{Bliokh2015}%
  \BibitemOpen
  \bibfield  {author} {\bibinfo {author} {\bibfnamefont {K.~Y.}\ \bibnamefont
  {Bliokh}}, \bibinfo {author} {\bibfnamefont {F.~J.}\ \bibnamefont
  {Rodr{\'{i}}guez-Fortu{\~{n}}o}}, \bibinfo {author} {\bibfnamefont
  {F.}~\bibnamefont {Nori}},\ and\ \bibinfo {author} {\bibfnamefont {A.~V.}\
  \bibnamefont {Zayats}},\ }\bibfield  {title} {\bibinfo {title} {{Spin--orbit
  interactions of light}},\ }\href@noop {} {\bibfield  {journal} {\bibinfo
  {journal} {Nature Photonics}\ }\textbf {\bibinfo {volume} {9}},\ \bibinfo
  {pages} {796} (\bibinfo {year} {2015})}\BibitemShut {NoStop}%
\bibitem [{\citenamefont {Kotlyar}\ \emph {et~al.}(2020)\citenamefont
  {Kotlyar}, \citenamefont {Nalimov}, \citenamefont {Kovalev}, \citenamefont
  {Porfirev},\ and\ \citenamefont {Stafeev}}]{Kotlyar2020}%
  \BibitemOpen
  \bibfield  {author} {\bibinfo {author} {\bibfnamefont {V.~V.}\ \bibnamefont
  {Kotlyar}}, \bibinfo {author} {\bibfnamefont {A.~G.}\ \bibnamefont
  {Nalimov}}, \bibinfo {author} {\bibfnamefont {A.~A.}\ \bibnamefont
  {Kovalev}}, \bibinfo {author} {\bibfnamefont {A.~P.}\ \bibnamefont
  {Porfirev}},\ and\ \bibinfo {author} {\bibfnamefont {S.~S.}\ \bibnamefont
  {Stafeev}},\ }\bibfield  {title} {\bibinfo {title} {Spin-orbit and orbit-spin
  conversion in the sharp focus of laser light: Theory and experiment},\ }\href
  {https://doi.org/10.1103/PhysRevA.102.033502} {\bibfield  {journal} {\bibinfo
   {journal} {Phys. Rev. A}\ }\textbf {\bibinfo {volume} {102}},\ \bibinfo
  {pages} {033502} (\bibinfo {year} {2020})}\BibitemShut {NoStop}%
\bibitem [{\citenamefont {Ganic}\ \emph {et~al.}(2003)\citenamefont {Ganic},
  \citenamefont {Gan},\ and\ \citenamefont {Gu}}]{Ganic2003}%
  \BibitemOpen
  \bibfield  {author} {\bibinfo {author} {\bibfnamefont {D.}~\bibnamefont
  {Ganic}}, \bibinfo {author} {\bibfnamefont {X.}~\bibnamefont {Gan}},\ and\
  \bibinfo {author} {\bibfnamefont {M.}~\bibnamefont {Gu}},\ }\bibfield
  {title} {\bibinfo {title} {Focusing of doughnut laser beams by a high
  numerical-aperture objective in free space},\ }\href
  {https://doi.org/10.1364/oe.11.002747} {\bibfield  {journal} {\bibinfo
  {journal} {Optics Express}\ }\textbf {\bibinfo {volume} {11}},\ \bibinfo
  {pages} {2747} (\bibinfo {year} {2003})}\BibitemShut {NoStop}%
\bibitem [{\citenamefont {Bokor}\ \emph {et~al.}(2005)\citenamefont {Bokor},
  \citenamefont {Iketaki}, \citenamefont {Watanabe},\ and\ \citenamefont
  {Fujii}}]{Bokor2005}%
  \BibitemOpen
  \bibfield  {author} {\bibinfo {author} {\bibfnamefont {N.}~\bibnamefont
  {Bokor}}, \bibinfo {author} {\bibfnamefont {Y.}~\bibnamefont {Iketaki}},
  \bibinfo {author} {\bibfnamefont {T.}~\bibnamefont {Watanabe}},\ and\
  \bibinfo {author} {\bibfnamefont {M.}~\bibnamefont {Fujii}},\ }\bibfield
  {title} {\bibinfo {title} {Investigation of polarization effects for
  high-numerical-aperture first-order laguerre-gaussian beams by 2d scanning
  with a single fluorescent microbead},\ }\href
  {https://doi.org/10.1364/opex.13.010440} {\bibfield  {journal} {\bibinfo
  {journal} {Optics Express}\ }\textbf {\bibinfo {volume} {13}},\ \bibinfo
  {pages} {10440} (\bibinfo {year} {2005})}\BibitemShut {NoStop}%
\bibitem [{\citenamefont {Iketaki}\ \emph {et~al.}(2007)\citenamefont
  {Iketaki}, \citenamefont {Watanabe}, \citenamefont {Bokor},\ and\
  \citenamefont {Fujii}}]{Iketaki2007}%
  \BibitemOpen
  \bibfield  {author} {\bibinfo {author} {\bibfnamefont {Y.}~\bibnamefont
  {Iketaki}}, \bibinfo {author} {\bibfnamefont {T.}~\bibnamefont {Watanabe}},
  \bibinfo {author} {\bibfnamefont {N.}~\bibnamefont {Bokor}},\ and\ \bibinfo
  {author} {\bibfnamefont {M.}~\bibnamefont {Fujii}},\ }\bibfield  {title}
  {\bibinfo {title} {Investigation of the center intensity of first- and
  second-order laguerre-gaussian beams with linear and circular polarization},\
  }\href {https://doi.org/10.1364/ol.32.002357} {\bibfield  {journal} {\bibinfo
   {journal} {Optics Letters}\ }\textbf {\bibinfo {volume} {32}},\ \bibinfo
  {pages} {2357} (\bibinfo {year} {2007})}\BibitemShut {NoStop}%
\bibitem [{\citenamefont {Monteiro}\ \emph {et~al.}(2009)\citenamefont
  {Monteiro}, \citenamefont {Neto},\ and\ \citenamefont
  {Nussenzveig}}]{Monteiro2009}%
  \BibitemOpen
  \bibfield  {author} {\bibinfo {author} {\bibfnamefont {P.~B.}\ \bibnamefont
  {Monteiro}}, \bibinfo {author} {\bibfnamefont {P.~A.~M.}\ \bibnamefont
  {Neto}},\ and\ \bibinfo {author} {\bibfnamefont {H.~M.}\ \bibnamefont
  {Nussenzveig}},\ }\bibfield  {title} {\bibinfo {title} {Angular momentum of
  focused beams: Beyond the paraxial approximation},\ }\href
  {https://doi.org/10.1103/PhysRevA.79.033830} {\bibfield  {journal} {\bibinfo
  {journal} {Phys. Rev. A}\ }\textbf {\bibinfo {volume} {79}},\ \bibinfo
  {pages} {033830} (\bibinfo {year} {2009})}\BibitemShut {NoStop}%
\bibitem [{\citenamefont {Philbin}(2018)}]{Philbin2018}%
  \BibitemOpen
  \bibfield  {author} {\bibinfo {author} {\bibfnamefont {T.~G.}\ \bibnamefont
  {Philbin}},\ }\bibfield  {title} {\bibinfo {title} {Some exact solutions for
  light beams},\ }\href {https://doi.org/10.1088/2040-8986/aade6d} {\bibfield
  {journal} {\bibinfo  {journal} {Journal of Optics}\ }\textbf {\bibinfo
  {volume} {20}},\ \bibinfo {pages} {105603} (\bibinfo {year}
  {2018})}\BibitemShut {NoStop}%
\bibitem [{\citenamefont {Kotlyar}\ \emph {et~al.}(2019)\citenamefont
  {Kotlyar}, \citenamefont {Stafeev},\ and\ \citenamefont
  {Nalimov}}]{Kotlyar2019}%
  \BibitemOpen
  \bibfield  {author} {\bibinfo {author} {\bibfnamefont {V.~V.}\ \bibnamefont
  {Kotlyar}}, \bibinfo {author} {\bibfnamefont {S.~S.}\ \bibnamefont
  {Stafeev}},\ and\ \bibinfo {author} {\bibfnamefont {A.~G.}\ \bibnamefont
  {Nalimov}},\ }\bibfield  {title} {\bibinfo {title} {Energy backflow in the
  focus of a light beam with phase or polarization singularity},\ }\href
  {https://doi.org/10.1103/PhysRevA.99.033840} {\bibfield  {journal} {\bibinfo
  {journal} {Phys. Rev. A}\ }\textbf {\bibinfo {volume} {99}},\ \bibinfo
  {pages} {033840} (\bibinfo {year} {2019})}\BibitemShut {NoStop}%
\bibitem [{\citenamefont {Volke-Sepulveda}\ \emph {et~al.}(2002)\citenamefont
  {Volke-Sepulveda}, \citenamefont {Garc\'es-Ch\'avez}, \citenamefont
  {Ch\'avez-Cerda}, \citenamefont {Arlt},\ and\ \citenamefont
  {Dholakia}}]{Volke-Sepulveda2002}%
  \BibitemOpen
  \bibfield  {author} {\bibinfo {author} {\bibfnamefont {K.}~\bibnamefont
  {Volke-Sepulveda}}, \bibinfo {author} {\bibfnamefont {V.}~\bibnamefont
  {Garc\'es-Ch\'avez}}, \bibinfo {author} {\bibfnamefont {S.}~\bibnamefont
  {Ch\'avez-Cerda}}, \bibinfo {author} {\bibfnamefont {J.}~\bibnamefont
  {Arlt}},\ and\ \bibinfo {author} {\bibfnamefont {K.}~\bibnamefont
  {Dholakia}},\ }\bibfield  {title} {\bibinfo {title} {Orbital angular momentum
  of a high-order {B}essel light beam},\ }\href
  {https://doi.org/10.1088/1464-4266/4/2/373} {\bibfield  {journal} {\bibinfo
  {journal} {Journal of Optics B: Quantum and Semiclassical Optics}\ }\textbf
  {\bibinfo {volume} {4}},\ \bibinfo {pages} {S82} (\bibinfo {year}
  {2002})}\BibitemShut {NoStop}%
\bibitem [{\citenamefont {Curtis}\ and\ \citenamefont
  {Grier}(2003)}]{Curtis2003}%
  \BibitemOpen
  \bibfield  {author} {\bibinfo {author} {\bibfnamefont {J.~E.}\ \bibnamefont
  {Curtis}}\ and\ \bibinfo {author} {\bibfnamefont {D.~G.}\ \bibnamefont
  {Grier}},\ }\bibfield  {title} {\bibinfo {title} {Structure of optical
  vortices},\ }\href {https://doi.org/10.1103/PhysRevLett.90.133901} {\bibfield
   {journal} {\bibinfo  {journal} {Phys. Rev. Lett.}\ }\textbf {\bibinfo
  {volume} {90}},\ \bibinfo {pages} {133901} (\bibinfo {year}
  {2003})}\BibitemShut {NoStop}%
\bibitem [{\citenamefont {Ng}\ \emph {et~al.}(2010)\citenamefont {Ng},
  \citenamefont {Lin},\ and\ \citenamefont {Chan}}]{Ng2010}%
  \BibitemOpen
  \bibfield  {author} {\bibinfo {author} {\bibfnamefont {J.}~\bibnamefont
  {Ng}}, \bibinfo {author} {\bibfnamefont {Z.}~\bibnamefont {Lin}},\ and\
  \bibinfo {author} {\bibfnamefont {C.~T.}\ \bibnamefont {Chan}},\ }\bibfield
  {title} {\bibinfo {title} {Theory of optical trapping by an optical vortex
  beam},\ }\href {https://doi.org/10.1103/PhysRevLett.104.103601} {\bibfield
  {journal} {\bibinfo  {journal} {Phys. Rev. Lett.}\ }\textbf {\bibinfo
  {volume} {104}},\ \bibinfo {pages} {103601} (\bibinfo {year}
  {2010})}\BibitemShut {NoStop}%
\bibitem [{\citenamefont {Zhou}\ \emph {et~al.}(2017)\citenamefont {Zhou},
  \citenamefont {Xiao}, \citenamefont {Chen},\ and\ \citenamefont
  {Zhao}}]{Zhou2017}%
  \BibitemOpen
  \bibfield  {author} {\bibinfo {author} {\bibfnamefont {L.-M.}\ \bibnamefont
  {Zhou}}, \bibinfo {author} {\bibfnamefont {K.-W.}\ \bibnamefont {Xiao}},
  \bibinfo {author} {\bibfnamefont {J.}~\bibnamefont {Chen}},\ and\ \bibinfo
  {author} {\bibfnamefont {N.}~\bibnamefont {Zhao}},\ }\bibfield  {title}
  {\bibinfo {title} {Optical levitation of nanodiamonds by doughnut beams in
  vacuum},\ }\href {https://doi.org/https://doi.org/10.1002/lpor.201600284}
  {\bibfield  {journal} {\bibinfo  {journal} {Laser \& Photonics Reviews}\
  }\textbf {\bibinfo {volume} {11}},\ \bibinfo {pages} {1600284} (\bibinfo
  {year} {2017})},\ \Eprint
  {https://arxiv.org/abs/https://onlinelibrary.wiley.com/doi/pdf/10.1002/lpor.201600284}
  {https://onlinelibrary.wiley.com/doi/pdf/10.1002/lpor.201600284} \BibitemShut
  {NoStop}%
\bibitem [{\citenamefont {T\"or\"ok}\ \emph {et~al.}(1995)\citenamefont
  {T\"or\"ok}, \citenamefont {Varga}, \citenamefont {Laczik},\ and\
  \citenamefont {Booker}}]{Torok95}%
  \BibitemOpen
  \bibfield  {author} {\bibinfo {author} {\bibfnamefont {P.}~\bibnamefont
  {T\"or\"ok}}, \bibinfo {author} {\bibfnamefont {P.}~\bibnamefont {Varga}},
  \bibinfo {author} {\bibfnamefont {Z.}~\bibnamefont {Laczik}},\ and\ \bibinfo
  {author} {\bibfnamefont {G.~R.}\ \bibnamefont {Booker}},\ }\bibfield  {title}
  {\bibinfo {title} {{Electromagnetic diffraction of light focused through a
  planar interface between materials of mismatched refractive indices: an
  integral representation}},\ }\href@noop {} {\bibfield  {journal} {\bibinfo
  {journal} {Journal of the Optical Society of America A}\ }\textbf {\bibinfo
  {volume} {12}},\ \bibinfo {pages} {325} (\bibinfo {year} {1995})}\BibitemShut
  {NoStop}%
\bibitem [{\citenamefont {Seifert}(2012)}]{Seifert2012}%
  \BibitemOpen
  \bibfield  {author} {\bibinfo {author} {\bibfnamefont {U.}~\bibnamefont
  {Seifert}},\ }\bibfield  {title} {\bibinfo {title} {Stochastic
  thermodynamics, fluctuation theorems and molecular machines},\ }\href
  {https://doi.org/10.1088/0034-4885/75/12/126001} {\bibfield  {journal}
  {\bibinfo  {journal} {Reports on Progress in Physics}\ }\textbf {\bibinfo
  {volume} {75}},\ \bibinfo {pages} {126001} (\bibinfo {year}
  {2012})}\BibitemShut {NoStop}%
\bibitem [{\citenamefont {Ciliberto}(2017)}]{Ciliberto2017}%
  \BibitemOpen
  \bibfield  {author} {\bibinfo {author} {\bibfnamefont {S.}~\bibnamefont
  {Ciliberto}},\ }\bibfield  {title} {\bibinfo {title} {Experiments in
  stochastic thermodynamics: Short history and perspectives},\ }\href
  {https://doi.org/10.1103/PhysRevX.7.021051} {\bibfield  {journal} {\bibinfo
  {journal} {Phys. Rev. X}\ }\textbf {\bibinfo {volume} {7}},\ \bibinfo {pages}
  {021051} (\bibinfo {year} {2017})}\BibitemShut {NoStop}%
\bibitem [{\citenamefont {Seifert}(2019)}]{Seifert2019}%
  \BibitemOpen
  \bibfield  {author} {\bibinfo {author} {\bibfnamefont {U.}~\bibnamefont
  {Seifert}},\ }\bibfield  {title} {\bibinfo {title} {From stochastic
  thermodynamics to thermodynamic inference},\ }\href
  {https://doi.org/10.1146/annurev-conmatphys-031218-013554} {\bibfield
  {journal} {\bibinfo  {journal} {Annual Review of Condensed Matter Physics}\
  }\textbf {\bibinfo {volume} {10}},\ \bibinfo {pages} {171} (\bibinfo {year}
  {2019})},\ \Eprint
  {https://arxiv.org/abs/https://doi.org/10.1146/annurev-conmatphys-031218-013554}
  {https://doi.org/10.1146/annurev-conmatphys-031218-013554} \BibitemShut
  {NoStop}%
\bibitem [{\citenamefont {Rold{\'a}n}\ \emph {et~al.}(2014)\citenamefont
  {Rold{\'a}n}, \citenamefont {Martinez}, \citenamefont {Parrondo},\ and\
  \citenamefont {Petrov}}]{Roldan2014}%
  \BibitemOpen
  \bibfield  {author} {\bibinfo {author} {\bibfnamefont {{\'E}.}~\bibnamefont
  {Rold{\'a}n}}, \bibinfo {author} {\bibfnamefont {I.~A.}\ \bibnamefont
  {Martinez}}, \bibinfo {author} {\bibfnamefont {J.~M.}\ \bibnamefont
  {Parrondo}},\ and\ \bibinfo {author} {\bibfnamefont {D.}~\bibnamefont
  {Petrov}},\ }\bibfield  {title} {\bibinfo {title} {Universal features in the
  energetics of symmetry breaking},\ }\href {https://doi.org/10.1038/nphys3230}
  {\bibfield  {journal} {\bibinfo  {journal} {Nature Physics}\ }\textbf
  {\bibinfo {volume} {10}},\ \bibinfo {pages} {457} (\bibinfo {year}
  {2014})}\BibitemShut {NoStop}%
\bibitem [{\citenamefont {Neto}\ and\ \citenamefont
  {Nussenzveig}(2000)}]{Neto2000}%
  \BibitemOpen
  \bibfield  {author} {\bibinfo {author} {\bibfnamefont {P.~A.~M.}\
  \bibnamefont {Neto}}\ and\ \bibinfo {author} {\bibfnamefont {H.~M.}\
  \bibnamefont {Nussenzveig}},\ }\bibfield  {title} {\bibinfo {title} {Theory
  of optical tweezers},\ }\href {https://doi.org/10.1209/epl/i2000-00327-4}
  {\bibfield  {journal} {\bibinfo  {journal} {Europhysics Letters ({EPL})}\
  }\textbf {\bibinfo {volume} {50}},\ \bibinfo {pages} {702} (\bibinfo {year}
  {2000})}\BibitemShut {NoStop}%
\bibitem [{\citenamefont {Mazolli}\ \emph {et~al.}(2003)\citenamefont
  {Mazolli}, \citenamefont {Neto},\ and\ \citenamefont
  {Nussenzveig}}]{Mazolli2003}%
  \BibitemOpen
  \bibfield  {author} {\bibinfo {author} {\bibfnamefont {A.}~\bibnamefont
  {Mazolli}}, \bibinfo {author} {\bibfnamefont {P.~A.~M.}\ \bibnamefont
  {Neto}},\ and\ \bibinfo {author} {\bibfnamefont {H.~M.}\ \bibnamefont
  {Nussenzveig}},\ }\bibfield  {title} {\bibinfo {title} {Theory of trapping
  forces in optical tweezers},\ }\href {https://doi.org/10.1098/rspa.2003.1164}
  {\bibfield  {journal} {\bibinfo  {journal} {Proceedings of the Royal Society
  of London. Series A: Mathematical, Physical and Engineering Sciences}\
  }\textbf {\bibinfo {volume} {459}},\ \bibinfo {pages} {3021} (\bibinfo {year}
  {2003})}\BibitemShut {NoStop}%
\bibitem [{\citenamefont {Viana}\ \emph {et~al.}(2007)\citenamefont {Viana},
  \citenamefont {Rocha}, \citenamefont {Mesquita}, \citenamefont {Mazolli},
  \citenamefont {Maia~Neto},\ and\ \citenamefont {Nussenzveig}}]{Viana2007}%
  \BibitemOpen
  \bibfield  {author} {\bibinfo {author} {\bibfnamefont {N.~B.}\ \bibnamefont
  {Viana}}, \bibinfo {author} {\bibfnamefont {M.~S.}\ \bibnamefont {Rocha}},
  \bibinfo {author} {\bibfnamefont {O.~N.}\ \bibnamefont {Mesquita}}, \bibinfo
  {author} {\bibfnamefont {A.}~\bibnamefont {Mazolli}}, \bibinfo {author}
  {\bibfnamefont {P.~A.}\ \bibnamefont {Maia~Neto}},\ and\ \bibinfo {author}
  {\bibfnamefont {H.~M.}\ \bibnamefont {Nussenzveig}},\ }\bibfield  {title}
  {\bibinfo {title} {Towards absolute calibration of optical tweezers},\ }\href
  {https://doi.org/10.1103/PhysRevE.75.021914} {\bibfield  {journal} {\bibinfo
  {journal} {Phys. Rev. E}\ }\textbf {\bibinfo {volume} {75}},\ \bibinfo
  {pages} {021914} (\bibinfo {year} {2007})}\BibitemShut {NoStop}%
\bibitem [{\citenamefont {Dutra}\ \emph {et~al.}(2007)\citenamefont {Dutra},
  \citenamefont {Viana}, \citenamefont {{Maia Neto}},\ and\ \citenamefont
  {Nussenzveig}}]{Dutra2007}%
  \BibitemOpen
  \bibfield  {author} {\bibinfo {author} {\bibfnamefont {R.~S.}\ \bibnamefont
  {Dutra}}, \bibinfo {author} {\bibfnamefont {N.~B.}\ \bibnamefont {Viana}},
  \bibinfo {author} {\bibfnamefont {P.~A.}\ \bibnamefont {{Maia Neto}}},\ and\
  \bibinfo {author} {\bibfnamefont {H.~M.}\ \bibnamefont {Nussenzveig}},\
  }\bibfield  {title} {\bibinfo {title} {{Polarization effects in optical
  tweezers}},\ }\href@noop {} {\bibfield  {journal} {\bibinfo  {journal}
  {Journal of Optics A: Pure and Applied Optics}\ }\textbf {\bibinfo {volume}
  {9}},\ \bibinfo {pages} {S221} (\bibinfo {year} {2007})}\BibitemShut
  {NoStop}%
\bibitem [{\citenamefont {Richards}\ and\ \citenamefont
  {Wolf}(1959)}]{Richards1959}%
  \BibitemOpen
  \bibfield  {author} {\bibinfo {author} {\bibfnamefont {B.}~\bibnamefont
  {Richards}}\ and\ \bibinfo {author} {\bibfnamefont {E.}~\bibnamefont
  {Wolf}},\ }\bibfield  {title} {\bibinfo {title} {Electromagnetic diffraction
  in optical systems, {II}. structure of the image field in an aplanatic
  system},\ }\href {https://doi.org/10.1098/rspa.1959.0200} {\bibfield
  {journal} {\bibinfo  {journal} {Proceedings of the Royal Society of London.
  Series A. Mathematical and Physical Sciences}\ }\textbf {\bibinfo {volume}
  {253}},\ \bibinfo {pages} {358} (\bibinfo {year} {1959})}\BibitemShut
  {NoStop}%
\bibitem [{\citenamefont {Rosales-Guzm{\'a}n}\ and\ \citenamefont
  {Forbes}(2017)}]{rosales2017shape}%
  \BibitemOpen
  \bibfield  {author} {\bibinfo {author} {\bibfnamefont {C.}~\bibnamefont
  {Rosales-Guzm{\'a}n}}\ and\ \bibinfo {author} {\bibfnamefont
  {A.}~\bibnamefont {Forbes}},\ }\href@noop {} {\emph {\bibinfo {title} {How to
  shape light with spatial light modulators}}}\ (\bibinfo  {publisher} {SPIE
  Press},\ \bibinfo {year} {2017})\BibitemShut {NoStop}%
\bibitem [{\citenamefont {Born}\ and\ \citenamefont {Wolf}(2019)}]{Born2019}%
  \BibitemOpen
  \bibfield  {author} {\bibinfo {author} {\bibfnamefont {M.}~\bibnamefont
  {Born}}\ and\ \bibinfo {author} {\bibfnamefont {E.}~\bibnamefont {Wolf}},\
  }\href@noop {} {\emph {\bibinfo {title} {Principles of Optics}}}\ (\bibinfo
  {publisher} {Cambridge University Press},\ \bibinfo {year} {2019})\
  Chap.~\bibinfo {chapter} {IX}\BibitemShut {NoStop}%
\bibitem [{\citenamefont {Dutra}\ \emph {et~al.}(2012)\citenamefont {Dutra},
  \citenamefont {Viana}, \citenamefont {Neto},\ and\ \citenamefont
  {Nussenzveig}}]{Dutra2012}%
  \BibitemOpen
  \bibfield  {author} {\bibinfo {author} {\bibfnamefont {R.~S.}\ \bibnamefont
  {Dutra}}, \bibinfo {author} {\bibfnamefont {N.~B.}\ \bibnamefont {Viana}},
  \bibinfo {author} {\bibfnamefont {P.~A.~M.}\ \bibnamefont {Neto}},\ and\
  \bibinfo {author} {\bibfnamefont {H.~M.}\ \bibnamefont {Nussenzveig}},\
  }\bibfield  {title} {\bibinfo {title} {Absolute calibration of optical
  tweezers including aberrations},\ }\href {https://doi.org/10.1063/1.3699273}
  {\bibfield  {journal} {\bibinfo  {journal} {Applied Physics Letters}\
  }\textbf {\bibinfo {volume} {100}},\ \bibinfo {pages} {131115} (\bibinfo
  {year} {2012})}\BibitemShut {NoStop}%
\bibitem [{\citenamefont {Dutra}\ \emph {et~al.}(2014)\citenamefont {Dutra},
  \citenamefont {Viana}, \citenamefont {Neto},\ and\ \citenamefont
  {Nussenzveig}}]{Dutra2014}%
  \BibitemOpen
  \bibfield  {author} {\bibinfo {author} {\bibfnamefont {R.~S.}\ \bibnamefont
  {Dutra}}, \bibinfo {author} {\bibfnamefont {N.~B.}\ \bibnamefont {Viana}},
  \bibinfo {author} {\bibfnamefont {P.~A.~M.}\ \bibnamefont {Neto}},\ and\
  \bibinfo {author} {\bibfnamefont {H.~M.}\ \bibnamefont {Nussenzveig}},\
  }\bibfield  {title} {\bibinfo {title} {Absolute calibration of forces in
  optical tweezers},\ }\href {https://doi.org/10.1103/physreva.90.013825}
  {\bibfield  {journal} {\bibinfo  {journal} {Physical Review A}\ }\textbf
  {\bibinfo {volume} {90}},\ \bibinfo {pages} {013825} (\bibinfo {year}
  {2014})}\BibitemShut {NoStop}%
\bibitem [{\citenamefont {Diniz}\ \emph {et~al.}(2019)\citenamefont {Diniz},
  \citenamefont {Dutra}, \citenamefont {Pires}, \citenamefont {Viana},
  \citenamefont {Nussenzveig},\ and\ \citenamefont {Neto}}]{Diniz2019}%
  \BibitemOpen
  \bibfield  {author} {\bibinfo {author} {\bibfnamefont {K.}~\bibnamefont
  {Diniz}}, \bibinfo {author} {\bibfnamefont {R.~S.}\ \bibnamefont {Dutra}},
  \bibinfo {author} {\bibfnamefont {L.~B.}\ \bibnamefont {Pires}}, \bibinfo
  {author} {\bibfnamefont {N.~B.}\ \bibnamefont {Viana}}, \bibinfo {author}
  {\bibfnamefont {H.~M.}\ \bibnamefont {Nussenzveig}},\ and\ \bibinfo {author}
  {\bibfnamefont {P.~A.~M.}\ \bibnamefont {Neto}},\ }\bibfield  {title}
  {\bibinfo {title} {Negative optical torque on a microsphere in optical
  tweezers},\ }\href {https://doi.org/10.1364/OE.27.005905} {\bibfield
  {journal} {\bibinfo  {journal} {Opt. Express}\ }\textbf {\bibinfo {volume}
  {27}},\ \bibinfo {pages} {5905} (\bibinfo {year} {2019})}\BibitemShut
  {NoStop}%
\bibitem [{\citenamefont {Dutra}\ \emph {et~al.}(2016)\citenamefont {Dutra},
  \citenamefont {Maia~Neto}, \citenamefont {Nussenzveig},\ and\ \citenamefont
  {Flyvbjerg}}]{Dutra2016}%
  \BibitemOpen
  \bibfield  {author} {\bibinfo {author} {\bibfnamefont {R.~S.}\ \bibnamefont
  {Dutra}}, \bibinfo {author} {\bibfnamefont {P.~A.}\ \bibnamefont
  {Maia~Neto}}, \bibinfo {author} {\bibfnamefont {H.~M.}\ \bibnamefont
  {Nussenzveig}},\ and\ \bibinfo {author} {\bibfnamefont {H.}~\bibnamefont
  {Flyvbjerg}},\ }\bibfield  {title} {\bibinfo {title} {Theory of
  optical-tweezers forces near a plane interface},\ }\href@noop {} {\bibfield
  {journal} {\bibinfo  {journal} {Phys. Rev. A}\ }\textbf {\bibinfo {volume}
  {94}},\ \bibinfo {pages} {053848} (\bibinfo {year} {2016})}\BibitemShut
  {NoStop}%
\bibitem [{\citenamefont {Edmonds}(1957)}]{Edmonds1957}%
  \BibitemOpen
  \bibfield  {author} {\bibinfo {author} {\bibfnamefont {A.~R.}\ \bibnamefont
  {Edmonds}},\ }\href@noop {} {\emph {\bibinfo {title} {Angular Momentum in
  Quantum Mechanics}}}\ (\bibinfo  {publisher} {Princeton University Press},\
  \bibinfo {year} {1957})\BibitemShut {NoStop}%
\bibitem [{\citenamefont {Ashkin}(1992)}]{Ashkin1992}%
  \BibitemOpen
  \bibfield  {author} {\bibinfo {author} {\bibfnamefont {A.}~\bibnamefont
  {Ashkin}},\ }\bibfield  {title} {\bibinfo {title} {Forces of a single-beam
  gradient laser trap on a dielectric sphere in the ray optics regime},\ }\href
  {https://doi.org/https://doi.org/10.1016/S0006-3495(92)81860-X} {\bibfield
  {journal} {\bibinfo  {journal} {Biophysical Journal}\ }\textbf {\bibinfo
  {volume} {61}},\ \bibinfo {pages} {569} (\bibinfo {year} {1992})}\BibitemShut
  {NoStop}%
\bibitem [{\citenamefont {Bohren}\ and\ \citenamefont
  {Huffman}(1998)}]{Bohren1998}%
  \BibitemOpen
  \bibfield  {author} {\bibinfo {author} {\bibfnamefont {C.~F.}\ \bibnamefont
  {Bohren}}\ and\ \bibinfo {author} {\bibfnamefont {D.~R.}\ \bibnamefont
  {Huffman}},\ }\href@noop {} {\emph {\bibinfo {title} {Absorption and
  Scattering of Light by Small Particles}}}\ (\bibinfo  {publisher} {Wiley},\
  \bibinfo {year} {1998})\BibitemShut {NoStop}%
\bibitem [{DLM(2022)}]{DLMF-10}%
  \BibitemOpen
  \href@noop {} {\bibinfo {title} {{NIST Digital Library of Mathematical
  Functions}}},\ \bibinfo {howpublished} {http://dlmf.nist.gov/10, release
  1.1.7 of 2022-10-15, edited by F. W. J. Olver, A. B. Olde Daalhuis, D. W.
  Lozier, B. I. Schneider, R. F. Boisvert, C. W. Clark, B. R. Miller, B. V.
  Saunders, H. S. Cohl, and M. A. McClain} (\bibinfo {year} {2022})\BibitemShut
  {NoStop}%
\bibitem{SM} See Supplemental Material at [URL will be inserted by publisher] for videos of 
the optically trapped microsphere for different focal heights. From left to right we increase
the spherical aberration by moving the sample chamber downwards. The microsphere radius is 
$1.5\,\mu\rm{m}$ and the topological charge is $\ell=8$. 
\bibitem [{\citenamefont {Li}\ \emph {et~al.}(2021{\natexlab{b}})\citenamefont
  {Li}, \citenamefont {Zhou},\ and\ \citenamefont {Zhao}}]{Li2021}%
  \BibitemOpen
  \bibfield  {author} {\bibinfo {author} {\bibfnamefont {Y.}~\bibnamefont
  {Li}}, \bibinfo {author} {\bibfnamefont {L.-M.}\ \bibnamefont {Zhou}},\ and\
  \bibinfo {author} {\bibfnamefont {N.}~\bibnamefont {Zhao}},\ }\bibfield
  {title} {\bibinfo {title} {Anomalous motion of a particle levitated by
  {L}aguerre--{G}aussian beams},\ }\href {https://doi.org/10.1364/OL.405696}
  {\bibfield  {journal} {\bibinfo  {journal} {Opt. Lett.}\ }\textbf {\bibinfo
  {volume} {46}},\ \bibinfo {pages} {106} (\bibinfo {year}
  {2021}{\natexlab{b}})}\BibitemShut {NoStop}%
\bibitem [{\citenamefont {Roichman}\ \emph {et~al.}(2008)\citenamefont
  {Roichman}, \citenamefont {Sun}, \citenamefont {Stolarski},\ and\
  \citenamefont {Grier}}]{Roichman2008}%
  \BibitemOpen
  \bibfield  {author} {\bibinfo {author} {\bibfnamefont {Y.}~\bibnamefont
  {Roichman}}, \bibinfo {author} {\bibfnamefont {B.}~\bibnamefont {Sun}},
  \bibinfo {author} {\bibfnamefont {A.}~\bibnamefont {Stolarski}},\ and\
  \bibinfo {author} {\bibfnamefont {D.~G.}\ \bibnamefont {Grier}},\ }\bibfield
  {title} {\bibinfo {title} {Influence of nonconservative optical forces on the
  dynamics of optically trapped colloidal spheres: The fountain of
  probability},\ }\href {https://doi.org/10.1103/PhysRevLett.101.128301}
  {\bibfield  {journal} {\bibinfo  {journal} {Phys. Rev. Lett.}\ }\textbf
  {\bibinfo {volume} {101}},\ \bibinfo {pages} {128301} (\bibinfo {year}
  {2008})}\BibitemShut {NoStop}%
\bibitem [{\citenamefont {Huang}\ \emph {et~al.}(2022)\citenamefont {Huang},
  \citenamefont {Wan}, \citenamefont {Zhou}, \citenamefont {Guo}, \citenamefont
  {Zhao}, \citenamefont {Chen}, \citenamefont {Ng},\ and\ \citenamefont
  {Du}}]{Huang2022}%
  \BibitemOpen
  \bibfield  {author} {\bibinfo {author} {\bibfnamefont {D.}~\bibnamefont
  {Huang}}, \bibinfo {author} {\bibfnamefont {P.}~\bibnamefont {Wan}}, \bibinfo
  {author} {\bibfnamefont {L.}~\bibnamefont {Zhou}}, \bibinfo {author}
  {\bibfnamefont {H.}~\bibnamefont {Guo}}, \bibinfo {author} {\bibfnamefont
  {R.}~\bibnamefont {Zhao}}, \bibinfo {author} {\bibfnamefont {J.}~\bibnamefont
  {Chen}}, \bibinfo {author} {\bibfnamefont {J.}~\bibnamefont {Ng}},\ and\
  \bibinfo {author} {\bibfnamefont {J.}~\bibnamefont {Du}},\ }\bibfield
  {title} {\bibinfo {title} {Optical trapping core formation and general
  trapping mechanism in single-beam optical tweezers},\ }\href
  {https://doi.org/10.1088/1367-2630/ac643a} {\bibfield  {journal} {\bibinfo
  {journal} {New Journal of Physics}\ }\textbf {\bibinfo {volume} {24}},\
  \bibinfo {pages} {043043} (\bibinfo {year} {2022})}\BibitemShut {NoStop}%
\bibitem [{\citenamefont {Roichman}\ \emph {et~al.}(2006)\citenamefont
  {Roichman}, \citenamefont {Waldron}, \citenamefont {Gardel},\ and\
  \citenamefont {Grier}}]{Roichman2006}%
  \BibitemOpen
  \bibfield  {author} {\bibinfo {author} {\bibfnamefont {Y.}~\bibnamefont
  {Roichman}}, \bibinfo {author} {\bibfnamefont {A.}~\bibnamefont {Waldron}},
  \bibinfo {author} {\bibfnamefont {E.}~\bibnamefont {Gardel}},\ and\ \bibinfo
  {author} {\bibfnamefont {D.~G.}\ \bibnamefont {Grier}},\ }\bibfield  {title}
  {\bibinfo {title} {{Optical traps with geometric aberrations}},\ }\href@noop
  {} {\bibfield  {journal} {\bibinfo  {journal} {Applied Optics}\ }\textbf
  {\bibinfo {volume} {45}},\ \bibinfo {pages} {3425} (\bibinfo {year}
  {2006})}\BibitemShut {NoStop}%
\bibitem [{\citenamefont {Mart{\'\i}nez}\ \emph {et~al.}(2016)\citenamefont
  {Mart{\'\i}nez}, \citenamefont {Petrosyan}, \citenamefont {Gu{\'e}ry-Odelin},
  \citenamefont {Trizac},\ and\ \citenamefont {Ciliberto}}]{Martinez2016}%
  \BibitemOpen
  \bibfield  {author} {\bibinfo {author} {\bibfnamefont {I.~A.}\ \bibnamefont
  {Mart{\'\i}nez}}, \bibinfo {author} {\bibfnamefont {A.}~\bibnamefont
  {Petrosyan}}, \bibinfo {author} {\bibfnamefont {D.}~\bibnamefont
  {Gu{\'e}ry-Odelin}}, \bibinfo {author} {\bibfnamefont {E.}~\bibnamefont
  {Trizac}},\ and\ \bibinfo {author} {\bibfnamefont {S.}~\bibnamefont
  {Ciliberto}},\ }\bibfield  {title} {\bibinfo {title} {Engineered swift
  equilibration of a brownian particle},\ }\href
  {https://doi.org/10.1038/nphys3758} {\bibfield  {journal} {\bibinfo
  {journal} {Nature Physics}\ }\textbf {\bibinfo {volume} {12}},\ \bibinfo
  {pages} {843} (\bibinfo {year} {2016})}\BibitemShut {NoStop}%
\bibitem [{\citenamefont {Raynal}\ \emph {et~al.}(2023)\citenamefont {Raynal},
  \citenamefont {de~Guillebon}, \citenamefont {Gu\'ery-Odelin}, \citenamefont
  {Trizac}, \citenamefont {Lauret},\ and\ \citenamefont {Rondin}}]{Raynal2023}%
  \BibitemOpen
  \bibfield  {author} {\bibinfo {author} {\bibfnamefont {D.}~\bibnamefont
  {Raynal}}, \bibinfo {author} {\bibfnamefont {T.}~\bibnamefont
  {de~Guillebon}}, \bibinfo {author} {\bibfnamefont {D.}~\bibnamefont
  {Gu\'ery-Odelin}}, \bibinfo {author} {\bibfnamefont {E.}~\bibnamefont
  {Trizac}}, \bibinfo {author} {\bibfnamefont {J.-S.}\ \bibnamefont {Lauret}},\
  and\ \bibinfo {author} {\bibfnamefont {L.}~\bibnamefont {Rondin}},\
  }\bibfield  {title} {\bibinfo {title} {Shortcuts to equilibrium with a
  levitated particle in the underdamped regime},\ }\href@noop {} {\bibfield
  {journal} {\bibinfo  {journal} {Phys. Rev. Lett.}\ }\textbf {\bibinfo
  {volume} {131}},\ \bibinfo {pages} {087101} (\bibinfo {year}
  {2023})}\BibitemShut {NoStop}%
\bibitem [{\citenamefont {Pires}\ \emph {et~al.}(2023)\citenamefont {Pires},
  \citenamefont {Goerlich}, \citenamefont {da~Fonseca}, \citenamefont
  {Debiossac}, \citenamefont {Hervieux}, \citenamefont {Manfredi},\ and\
  \citenamefont {Genet}}]{Pires2023}%
  \BibitemOpen
  \bibfield  {author} {\bibinfo {author} {\bibfnamefont {L.~B.}\ \bibnamefont
  {Pires}}, \bibinfo {author} {\bibfnamefont {R.}~\bibnamefont {Goerlich}},
  \bibinfo {author} {\bibfnamefont {A.~L.}\ \bibnamefont {da~Fonseca}},
  \bibinfo {author} {\bibfnamefont {M.}~\bibnamefont {Debiossac}}, \bibinfo
  {author} {\bibfnamefont {P.-A.}\ \bibnamefont {Hervieux}}, \bibinfo {author}
  {\bibfnamefont {G.}~\bibnamefont {Manfredi}},\ and\ \bibinfo {author}
  {\bibfnamefont {C.}~\bibnamefont {Genet}},\ }\href
  {https://doi.org/10.48550/ARXIV.2302.06003} {\bibinfo {title} {Optimal
  time-entropy bounds and speed limits for brownian thermal shortcuts}}
  (\bibinfo {year} {2023}),\ \Eprint
  {https://arxiv.org/abs/https://arxiv.org/abs/2302.06003}
  {https://arxiv.org/abs/2302.06003} \BibitemShut {NoStop}%
\bibitem [{\citenamefont {Viana}\ \emph {et~al.}(2006)\citenamefont {Viana},
  \citenamefont {Rocha}, \citenamefont {Mesquita}, \citenamefont {Mazolli},\
  and\ \citenamefont {Neto}}]{Viana2006}%
  \BibitemOpen
  \bibfield  {author} {\bibinfo {author} {\bibfnamefont {N.~B.}\ \bibnamefont
  {Viana}}, \bibinfo {author} {\bibfnamefont {M.~S.}\ \bibnamefont {Rocha}},
  \bibinfo {author} {\bibfnamefont {O.~N.}\ \bibnamefont {Mesquita}}, \bibinfo
  {author} {\bibfnamefont {A.}~\bibnamefont {Mazolli}},\ and\ \bibinfo {author}
  {\bibfnamefont {P.~A.~M.}\ \bibnamefont {Neto}},\ }\bibfield  {title}
  {\bibinfo {title} {Characterization of objective transmittance for optical
  tweezers},\ }\href {https://doi.org/10.1364/ao.45.004263} {\bibfield
  {journal} {\bibinfo  {journal} {Applied Optics}\ }\textbf {\bibinfo {volume}
  {45}},\ \bibinfo {pages} {4263} (\bibinfo {year} {2006})}\BibitemShut
  {NoStop}%
\bibitem [{\citenamefont {Feitosa}\ and\ \citenamefont
  {Mesquita}(1991)}]{Feitosa1991}%
  \BibitemOpen
  \bibfield  {author} {\bibinfo {author} {\bibfnamefont {M.~I.~M.}\
  \bibnamefont {Feitosa}}\ and\ \bibinfo {author} {\bibfnamefont {O.~N.}\
  \bibnamefont {Mesquita}},\ }\bibfield  {title} {\bibinfo {title} {Wall-drag
  effect on diffusion of colloidal particles near surfaces: A photon
  correlation study},\ }\href {https://doi.org/10.1103/physreva.44.6677}
  {\bibfield  {journal} {\bibinfo  {journal} {Physical Review A}\ }\textbf
  {\bibinfo {volume} {44}},\ \bibinfo {pages} {6677} (\bibinfo {year}
  {1991})}\BibitemShut {NoStop}%
\bibitem [{\citenamefont {Sch\"affer}\ \emph {et~al.}(2007)\citenamefont
  {Sch\"affer}, \citenamefont {N{\o}rrelykke},\ and\ \citenamefont
  {Howard}}]{Schaffer2007}%
  \BibitemOpen
  \bibfield  {author} {\bibinfo {author} {\bibfnamefont {E.}~\bibnamefont
  {Sch\"affer}}, \bibinfo {author} {\bibfnamefont {S.~F.}\ \bibnamefont
  {N{\o}rrelykke}},\ and\ \bibinfo {author} {\bibfnamefont {J.}~\bibnamefont
  {Howard}},\ }\bibfield  {title} {\bibinfo {title} {Surface forces and drag
  coefficients of microspheres near a plane surface measured with optical
  tweezers},\ }\href@noop {} {\bibfield  {journal} {\bibinfo  {journal}
  {Langmuir}\ }\textbf {\bibinfo {volume} {23}},\ \bibinfo {pages} {3654}
  (\bibinfo {year} {2007})}\BibitemShut {NoStop}%
\end{thebibliography}
\end{document}